\documentclass[twocolumn,prl,aps,superscriptaddress]{revtex4-1}
\usepackage[latin9]{inputenc}
\usepackage{color}
\usepackage{amstext}
\usepackage{amssymb,cleveref}
\usepackage{graphicx}
\usepackage[unicode=true,pdfusetitle,
 bookmarks=false,
 breaklinks=false,pdfborder={0 0 1},backref=false,colorlinks=false]
 {hyperref}
\hypersetup{
 bookmarksnumbered=false,bookmarksopen=false}
\usepackage{cleveref}

\makeatletter

\@ifundefined{textcolor}{}
{
 \definecolor{BLACK}{gray}{0}
 \definecolor{WHITE}{gray}{1}
 \definecolor{RED}{rgb}{1,0,0}
 \definecolor{GREEN}{rgb}{0,1,0}
 \definecolor{BLUE}{rgb}{0,0,1}
 \definecolor{CYAN}{cmyk}{1,0,0,0}
 \definecolor{MAGENTA}{cmyk}{0,1,0,0}
 \definecolor{YELLOW}{cmyk}{0,0,1,0}
}

\usepackage{amsmath}
\usepackage{graphicx}
\usepackage{amssymb}
\usepackage{txfonts}
\usepackage{dsfont}

\makeatother

\begin{document}

\title{Pulsed Dynamical Decoupling for Fast and Robust Two-Qubit Gates on Trapped Ions}

\author{I. Arrazola}
\affiliation{Department of Physical Chemistry, University of the Basque Country UPV/EHU, Apartado 644, 48080 Bilbao, Spain}
\author{J. Casanova}
\affiliation{Institute for Theoretical Physics and IQST, Albert-Einstein-Allee 11, Universit\"at Ulm, D-89069 Ulm, Germany}
\author{J. S. Pedernales}
\affiliation{Institute for Theoretical Physics and IQST, Albert-Einstein-Allee 11, Universit\"at Ulm, D-89069 Ulm, Germany}
\author{Z.-Y. Wang}
\affiliation{Institute for Theoretical Physics and IQST, Albert-Einstein-Allee 11, Universit\"at Ulm, D-89069 Ulm, Germany}
\author{E. Solano}
\affiliation{Department of Physical Chemistry, University of the Basque Country UPV/EHU, Apartado 644, 48080 Bilbao, Spain}
\affiliation{IKERBASQUE,  Basque  Foundation  for  Science,  Maria  Diaz  de  Haro  3,  48013  Bilbao,  Spain}
\author{M. B.  Plenio}
\affiliation{Institute for Theoretical Physics and IQST, Albert-Einstein-Allee 11, Universit\"at Ulm, D-89069 Ulm, Germany}

\begin{abstract}
We propose a pulsed dynamical decoupling protocol as the generator of tunable, fast, and robust quantum phase 
gates between two microwave-driven trapped ion hyperfine qubits. The protocol consists of sequences of 
$\pi$-pulses acting on ions that are oriented along an externally applied magnetic field
gradient. In contrast to existing approaches, in our design the two vibrational modes of the ion chain cooperate
under the influence of the external microwave driving to achieve significantly increased gate speeds. Our scheme is robust against the dominant noise sources, which are errors on the magnetic field and microwave pulse 
intensities, as well as motional heating, predicting two-qubit gates with fidelities above $99.9\%$ in tens of microseconds. 

\end{abstract}
\maketitle

\section{Introduction}
Entangling quantum gates, faster than decoherence rates and with high accuracy, are crucial to 
quantum technologies~\cite{Nielsen}. Among the latter, trapped-ion systems~\cite{Leibfried03, Haffner08} 
are one of the most promising candidate platforms for the implementation of quantum computing 
and simulations \cite{Casanova12,Mezzacapo12,Pedernales14, DiCandia15, Arrazola16} in a systematic manner. Laser control 
techniques have so far provided the best entangling gates in ion traps, reaching two-qubit gate 
fidelities around ~$99.9\%$ at gate times between 30-100 $\mu s$~\cite{Ballance16,Gaebler16}. 
However, scaling such systems implies daunting technological challenges in setting up and 
controlling multiple laser sources.

In the last years, an alternative route that relies on the microwave quantum control of trapped ions has 
been proposed~\cite{Mintert01} and pursued in several laboratories~\cite{Weidt15, Piltz16}. Microwave 
control elements are manipulated entirely with electronic methods, enjoy greater stability than lasers, 
and are sufficiently small to be integrated in the trap electrodes~\cite{Harty16}. Furthermore, 
microwave control avoids the use of optical transitions whose spontaneous decay rate limits 
the achievable quantum gate speeds especially at high target fidelities~\cite{Plenio97}.

Two approaches to microwave control of trapped ions are usually considered~\cite{Wolk16}. These are the 
cases of far-field microwave radiation with a static magnetic field gradient~\cite{Mintert01, Khromova12}, 
and the near-field of microwave radiation~\cite{Ospelkaus08,Ospelkaus11}. If the qubit states are 
magnetically sensitive, it is crucial to use dynamical decoupling (DD) techniques for the stabilisation
of quantum gates~\cite{Timoney11, Bermudez12, Jonathan00, Puebla16, Puebla17}. Following this method, 
encouraging experimental results have been achieved for the far- and the near-field approaches, reaching 
two-qubit gate fidelities of $98.5\%$~\cite{Weidt16} and  $99.7\%$~\cite{Harty16} respectively for 
gate times in the millisecond range, i.e. several orders of magnitude longer than the oscillation 
period of the ion chain. This is because, both in laser- or microwave-based schemes, qubit-qubit interactions 
are mediated by a single motional mode which needs to be spectroscopically discriminated from the 
rest~\cite{Sorensen99, Sorensen00, Cohen15}. To guarantee this, the qubit-motion coupling should be much 
smaller than the detuning from the modes that are to be neglected, which imposes severe limitations 
on the speed of the resulting gate. Therefore, gates acting on time scales comparable with the oscillation 
period of the ions or faster, require necessarily the involvement of all vibrational modes. Building
on earlier theoretical work \cite{Duan04} the use of multiple modes has recently been explored experimentally 
for laser based systems~\cite{Mizrahi13}.
\begin{figure}[t!]
\centering
\includegraphics[width=0.8\linewidth]{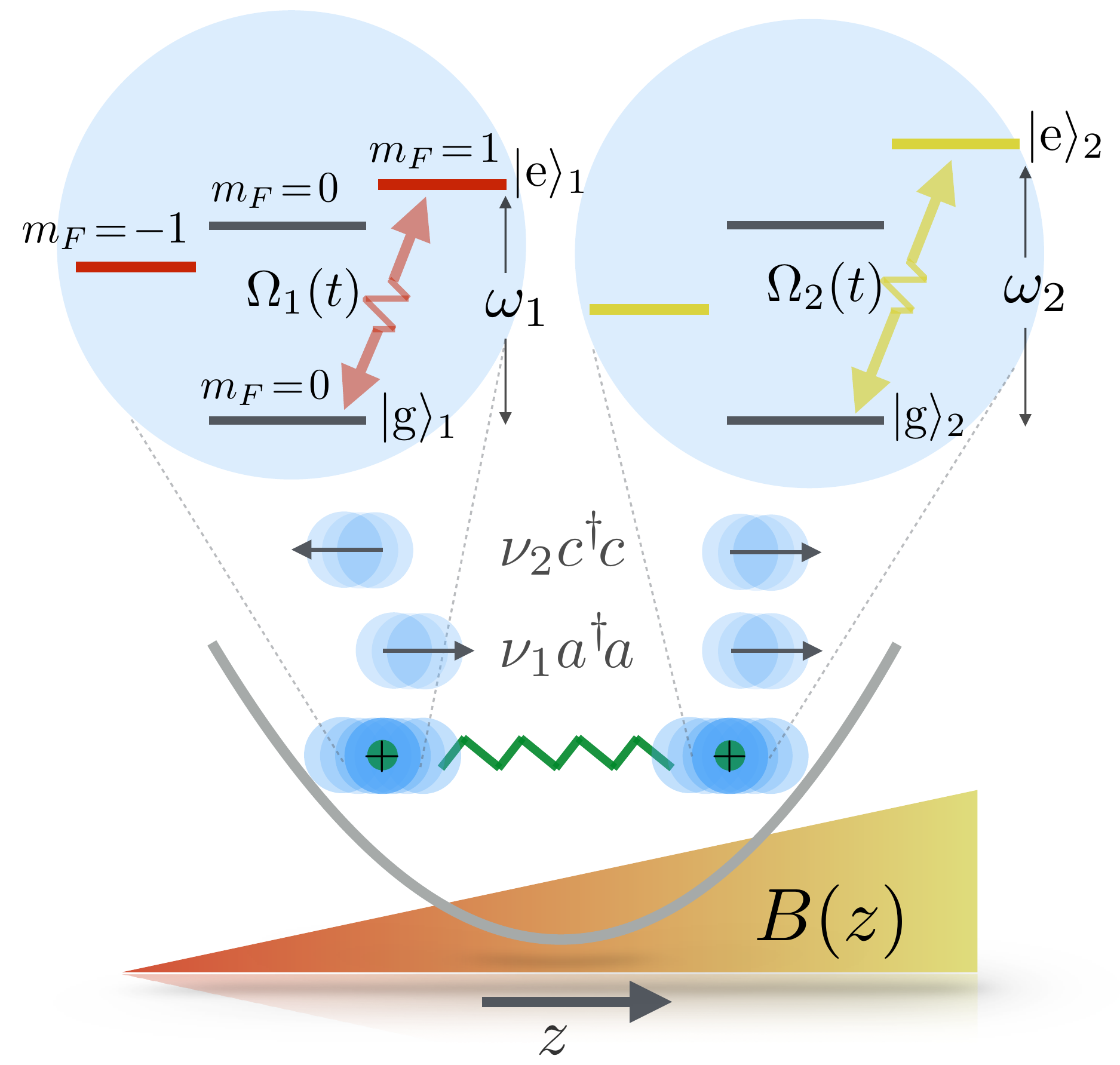}
\caption{Hyperfine levels of two $^{171}$Yb$^+$ ions. The magnetic field $B(z)$ removes the degeneracy of the $F=1$ manifold separating the $\{F=1, m_F=\pm1\}$ and $\{F=1, m_F=0\}$ levels of both ions by an amount of $\pm\gamma B(z_j)$ respectively. }\label{fig:Fig1}
\end{figure}

In this article, we propose a scheme leading to fast and high-fidelity two-qubit gates through a specifically 
designed sequence of microwave $\pi$-pulses acting in the presence of a magnetic field gradient. Our 
method employs the two vibrational modes in the axial direction of the two-ion chain leading to gate times 
approaching the inverse of the trap frequency. On top of that, the sequence is designed to protect 
qubits from uncontrolled noise sources. The high speed and robustness of our scheme results in two-qubit 
gates of high fidelity even in presence of motional heating. Our detailed numerical simulations show that 
state-of-the-art in microwave trapped-ion technology allows for two-qubit gates sufficiently fast to 
pave the way for scalable quantum computers.

Our work is organised as follows: In section II, we present the target system, i.e. a set of hyperfine trapped ions, and the method when working under ideal conditions. In section III, we introduce the microwave sequence we will use to achieve the proposed two-qubit gates. Section IV demonstrates the performance of our method when realistic experimental conditions are included in the model.

\section{The System}

We consider  two $^{171}$Yb$^+$ ions in a microwave quantum computer module~\cite{Lekitsch17}. The ions sit next to each other along the $z$ direction, where their coupled motion is described by the center-of-mass (com) and breathing modes, with frequencies ${\nu_1=\nu}$ and ${\nu_2=\sqrt{3}\nu}$ respectively~\cite{James98}. For each ion we define a quantum bit (qubit) using the states ${|{\rm g}\rangle\equiv\{F=0, m_F=0\}}$ 
and the magnetic sensitive ${| {\rm e} \rangle\equiv\{F=1, m_F=1\}}$ in the hyperfine manifold, see Fig.~\ref{fig:Fig1}. The presence of a magnetic field gradient will serve both to make the qubit frequencies $\omega_j$ different to each other, and to couple motion and qubit. The Hamiltonian of such a system is given by ($\hbar=1$)
\begin{eqnarray}\label{Hamiltonianbare}
    \nonumber H&=& \frac{\omega_1}{2}\sigma_1^z + \Delta_1 (b+b^\dag) \sigma_1^z - \Delta_2(c+c^\dag)\sigma_1^z\\
     &+& \frac{\omega_2}{2}\sigma_2^z + \Delta_1 (b+b^\dag) \sigma_2^z + \Delta_2(c+c^\dag)\sigma_2^z\\
    &+& \nu_1 b^\dag b + \nu_2 c^\dag c\nonumber,
\end{eqnarray}
where $b(b^\dag)$ and $c(c^\dag)$ are the bosonic annihilation(creation) operators associated to the com and breathing modes respectively, and the qubit-mode coupling is given by $\Delta_m= \frac{\gamma_e g_B}{8} \sqrt{\frac{\hbar}{M \nu_m}}$, where $\gamma_e \simeq (2\pi)\times 2.8$ MHz/G is the electronic gyromagnetic ratio and $M$ is the mass of each ion. We consider a linearly growing magnetic field which gradient $\partial B/\partial z = g_B$ is related to the energy difference between both ion-qubits as $\omega_2 - \omega_1=\gamma g_B \Delta z$, with $\Delta z$ the distance between the ions equilibrium positions. Detailed derivation of Eq.~(\ref{Hamiltonianbare}), along with the necessary conditions to neglect transitions to other hyperfine levels and coupling to radial modes are in appendices A and B. To control the performance of the two qubit gate, a bichromatic microwave field of frequencies $\omega_j$ and phases $\phi_j$ acts on the system with Rabi frequencies $\Omega_j$ and is described by the Hamiltonian $H_c(t)=\sum_{j=1}^2\Omega_j(t) (\sigma_1^x + \sigma_2^x)\cos(\omega_j t - \phi_j)$. 
Under such microwave control, Eq.~(\ref{Hamiltonianbare}) in a rotating frame w.r.t. $H_0= 
\nu_1 b^\dag b + \nu_2 c^\dag c + \frac{\omega_1}{2}\sigma^z_1 + \frac{\omega_2}{2}\sigma_2^z$ reads
\begin{eqnarray}\label{casi}
 H^{\rm I}(t)&=& \Delta_1(b e^{-i \nu_1 t} + b^\dag e^{i\nu_1 t}) \sigma_1^z - \Delta_2(ce^{-i\nu_2 t} + c^\dag e^{i \nu_2 t}) \sigma_1^z\\
\nonumber &+& \Delta_1(b e^{-i \nu_1 t} + b^\dag e^{i\nu_1 t}) \sigma_2^z + \Delta_2(ce^{-i\nu_2 t} + c^\dag e^{i \nu_2 t}) \sigma_2^z\\
\nonumber &+& \frac{\Omega_1(t)}{2}(\sigma_1^+ e^{i \phi_1} + \sigma_1^- e^{-i\phi_1}) + \frac{\Omega_2(t)}{2}(\sigma_2^+ e^{i \phi_2} + \sigma_2^- e^{-i\phi_2}).
\end{eqnarray}
In the above expression the non-resonant components of the microwave driving have been eliminated, see Appendix C for details. Here, off-resonant microwave components that rotate at  $\sim2\omega_j$ (tens of gigahertz for the 
$^{171}$Yb$^{+}$~\cite{Olmschenk07}) can be safely neglected invoking the RWA. However, those precessing  at $\sim |\omega_2- \omega_1|$ (tens of MHz for our simulated conditions) lead to significant undesired contributions whose cancelation is discussed in Sec.~\ref{SecSeq}.

Now we move to a rotating frame w.r.t. $\frac{\Omega_1(t)}{2}(\sigma_1^+ e^{i \phi_1} + \sigma_1^- e^{-i\phi_1}) + \frac{\Omega_2(t)}{2}(\sigma_2^+ e^{i \phi_2} + \sigma_2^- e^{-i\phi_2})$. The Rabi frequencies $\Omega_{1,2}(t)$ will be switched on and off, i.e. the driving is applied stroboscopically in the form of $\pi$-pulses, leading to
\begin{eqnarray}
\label{TheHamiltonian}
\nonumber H^{\rm II}(t)&=& f_{1}(t) \sigma_1^z[\Delta_1 b e^{-i \nu_1 t} -  \Delta_2ce^{-i\nu_2 t} + {\rm H.c.}] \\
 &+& f_{2}(t)\sigma_2^z[\Delta_1 b e^{-i \nu_1 t} + \Delta_2ce^{-i\nu_2 t} + {\rm H.c.}],
\end{eqnarray}
where the modulation functions $f_{j}(t)$ take the values $\pm 1$ depending on the number of  $\pi$-pulses applied to the $j$-th ion. More specifically, for an even (odd) number of pulses we have $f_{j} =1 (-1)$. The idealised description in Eq.~(\ref{TheHamiltonian}) assumes instantaneous $\pi$-pulses, which is a good approximation if the Rabi frequencies are much larger than any other frequency ($\Delta_{1,2}$ and $\nu_{1,2}$) in Eq.~(\ref{TheHamiltonian}). Nevertheless, to match realistic experimental conditions, our numerical simulations will consider sequences of finite $\pi$-pulses in the form of top-hat functions of lenght $t_{\pi} = \frac{\pi}{\Omega}$.

\begin{figure}[t]
\centering
\includegraphics[width=1\linewidth]{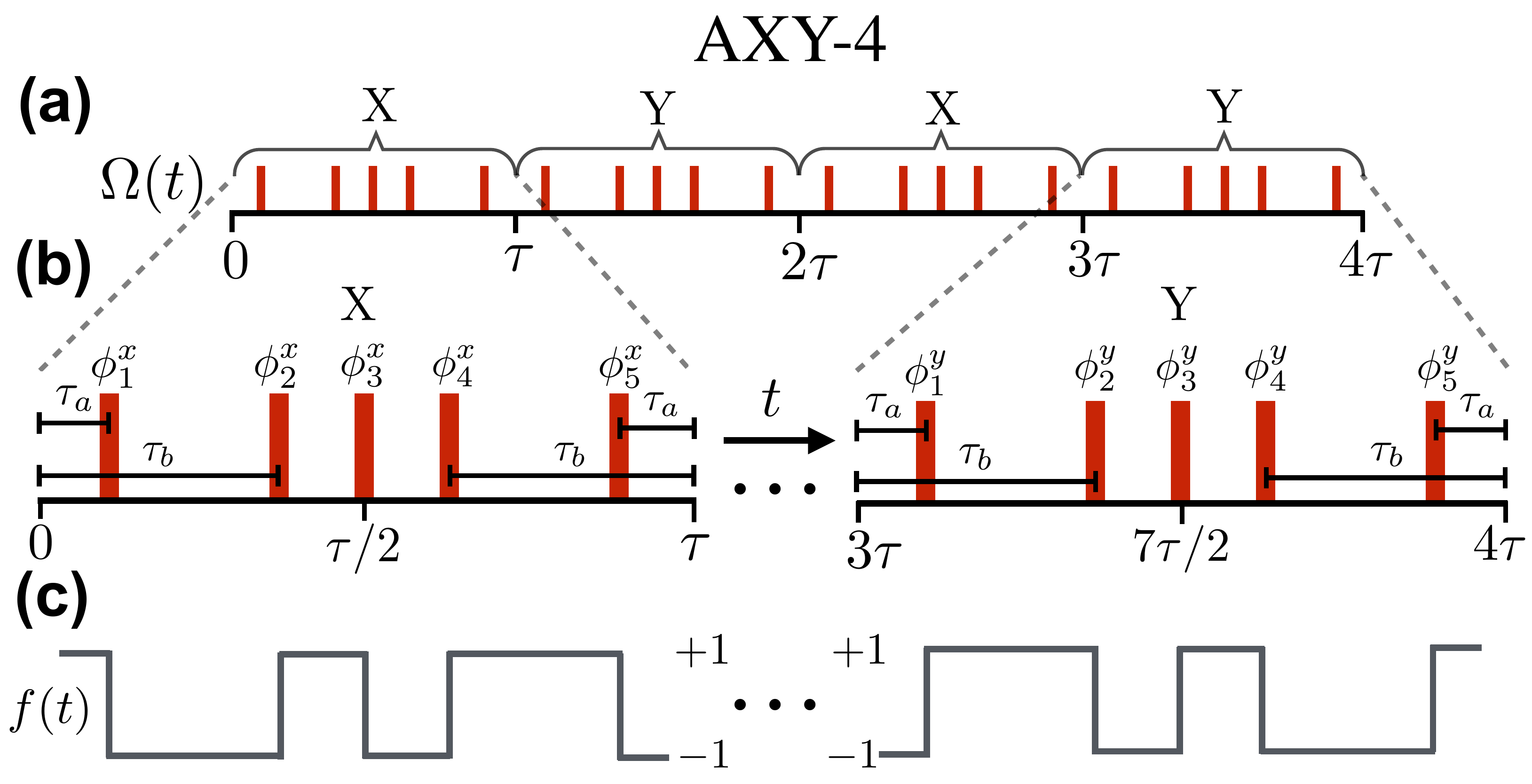}
\caption{{\bf(a)}  AXY-4 pulse sequence. Each composite pulse includes $5$ $\pi$-pulses with tunable distances between them. {\bf (b)} Zoom on the composite X and Y pulses with the corresponding pulse-phases in $H_c(t)$. {\bf (c)} Modulation function associated to the composite pulses.}\label{AXYblock}
\end{figure}

\begin{figure*}[t]
\centering
\includegraphics[width=1\linewidth]{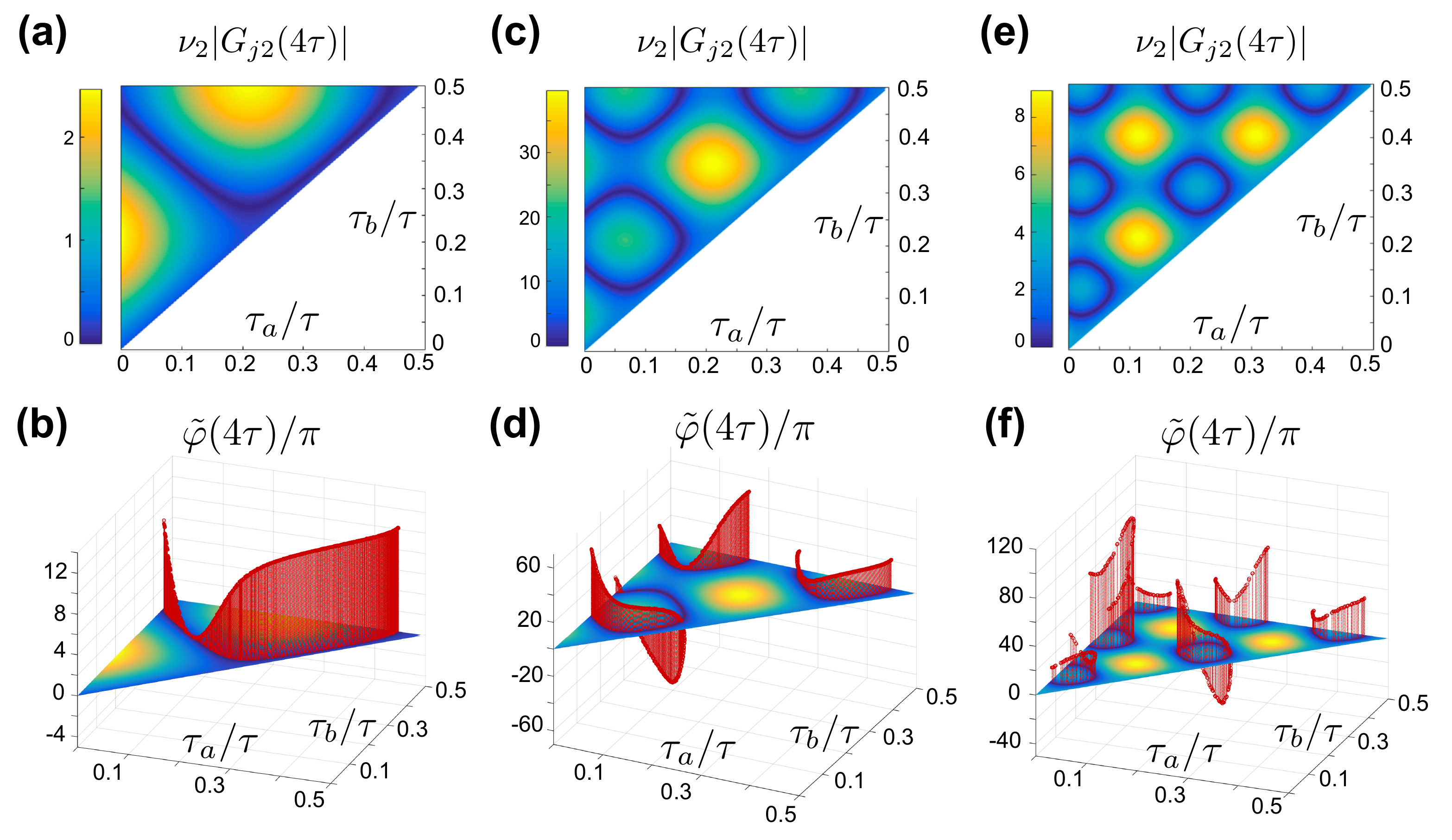}
\caption{Value of $G_{j2}(t)$ after an AXY-$4$ sequence as a function of $\tau_a$ and $\tau_b$ ($\tau_a<\tau_b<\tau/2$), for {\bf(a)}: $\tau=1\times2\pi/\nu_1$, {\bf(c)}: $\tau=2\times2\pi/\nu_1$, {\bf(e)}: $\tau=3\times2\pi/\nu_1$. The dark blue regions show the $\tau_a$ and $\tau_b$ values that correspond to a complete decoupling of the qubits with the modes at the end of the sequence. The  phases $\tilde{\varphi}(t)$ are represented in {\bf(b)},{\bf(d)},{\bf(f)} by the red panels.}
\label{Gplot}
\end{figure*}

The Schr\"odinger equation corresponding to Eq.~(\ref{TheHamiltonian})  is analytically solvable and leads to the propagator $U(t)=U_s(t) U_c(t)$ where 
\begin{equation}\label{solution1}
U_s(t)= \exp{\left[-i \sum_{j=1}^{2} \{\Delta_1 b \ G_{j1}(t) +(-1)^j \Delta_2 c \ G_{j2}(t)+ {\rm H.c.}\}\sigma_j^z\right]},
\end{equation}
and $U_c(t)~=\exp \left[i \varphi(t) \sigma_1^z \sigma_2^z\right]$, see Appendix D for derivation. The $G_{jm}(t)$ functions in $U_s(t)$ are $G_{jm}(t)= \int_0^t dt' f_j(t') e^{-i\nu_m t'}$, while the achieved two-qubit phase $\varphi(t)$ in $U_c(t)$ is
\begin{eqnarray}\label{phase}
\varphi(t)=\Bigg(\frac{\Delta_1}{\nu}\Bigg)^2[ \tilde{\varphi}_1(t)-\frac{1}{3\!\sqrt{3}} \tilde{\varphi}_2(t)]=\Bigg(\frac{\Delta_1}{\nu}\Bigg)^2 \tilde{\varphi}(t),
\end{eqnarray}
where $\tilde{\varphi}_m(t)=\nu_m^2 \ \! \Im \ \!\! {\int_{0}^t \!\!\! \ dt'} \big[ f_1(t')G_{2m}(t')+f_2(t')G_{1m}(t')  \big] \ e^{i\nu_m t'}$, and $\Im$ being the imaginary part of the subsequent integral. One can demonstrate that, at the end of the sequence,  $\tilde{\varphi}(t)$ does not depend on the values of $\Delta_{1,2}$ and $\nu_{1,2}$ but on the ratio between mode frequencies $\nu_2/\nu_1=\sqrt{3}$ (Appendix E). Hence, the study of $\tilde{\varphi}(t)$ covers all situations regardless of the value of $\Delta_{1,2}$ and $\nu_{1,2}$.

From the solution $U(t)$, it is clear that a $\pi$-pulse sequence of duration $t_{\rm gate}$, satisfying conditions
\begin{eqnarray}
\label{conditions}
G_{jm}(t_{\rm gate})=0, \  \  \varphi(t_{\rm gate})\neq 0,
\end{eqnarray}
results in a phase gate between the two qubits and leaves the hyperfine levels of the ions decoupled from their motion. To accomplish these two conditions, we will design a specific microwave pulse sequence that, in addition, will eliminate the dephasing noise due to magnetic field fluctuations or frequency offsets on the registers. Note that, if the latter are not averaged out, they would spoil the generation of a high-fidelity two-qubit gate.

\section{The Microwave Sequence}\label{SecMicro}

In order to satisfy Eqs.~(\ref{conditions}) we propose to use variations of the adaptive XY-n (AXY-n) 
family of decoupling sequences introduced in Ref.~\cite{Casanova15} for nanoscale nuclear magnetic 
resonance~\cite{Wu16, Wang16, Casanova16, Wang17, Casanova17}. Unlike previously used pulsed ion trap 
DD schemes~\cite{Piltz13}, AXY-n consists of n blocks of 5 non-equally separated $\pi$-pulses, as depicted 
in Fig. 2 for the AXY-4 case, where the interpulse spacing can be arbitrarily tuned while the sequence remains 
robust~\cite{Casanova15}. Each $\pi$-pulse is applied along an axis in the $x$-$y$ plane of the Bloch sphere of each qubit state that is rotated an angle $\phi$ (corresponding to $\phi_{1,2}$ in $H_c(t)$) w.r.t. the $x$ axis.

We define two blocks: the X block, made of 5 $\pi$-pulses along the axes corresponding to ${\vec\phi^x\equiv\{\phi^x_1,\phi^x_2,\phi^x_3,\phi^x_4,\phi^x_5\}=\{ \frac{\pi}{6}, \frac{\pi}{2}, 0 ,\frac{\pi}{2},\frac{\pi}{6}\}} + \zeta$, with $\zeta$ an arbitrary constant phase, and the Y block, with rotations along the same axes but shifted by a  $\pi/2$ phase, i.e. ${\vec{\phi}^y=\{ \frac{\pi}{6} + \frac{\pi}{2}, \pi, \frac{\pi}{2} ,\pi,\frac{\pi}{6} + \frac{\pi}{2}\}} + \zeta$. The sequence then has $n$ consecutive X and Y blocks with the same, tunable, interpulse spacing. For example, the AXY-$4$ sequence is XYXY. As illustrated in Fig.~\ref{AXYblock}{\bf (b)}, each block is symmetric and has a duration $\tau$. Therefore, within a five pulse block the time of application of the first and second pulses, $\tau_a$ and $\tau_b$ where $\tau_a<\tau_b<\tau/2$, together with $\tau$ define the whole sequence.

At the end of any AXY-n sequence of length $n\tau$, where $n$ is an even integer, the function $G_{jm}(n\tau)$ is zero for values of $\tau$ that are a multiple of the oscillation period of mode $m$, that is for $\nu_m \tau=2\pi r$ with $r\in \mathbb{N}$. This is due to the translational symmetry of the $f_{j}(t)$ functions, for which $f_j(t'+\tau)=-f_j(t')$ and $f_j(t'+2\tau)=f_j(t')$ holds, meaning that 
\begin{eqnarray}
G_{jm}(n\tau) &=&\int_{0}^{n\tau}dt^{\prime}f_{j}(t^{\prime})e^{-i\nu_{m}t^{\prime}}  \\
&\sum_{p=0}^{n/2-1}&\int_{0}^{\tau} f_j(t')\Big(e^{-i\nu_m[t'+2p\tau]}-e^{-i\nu_m[t'+(2p+1)\tau]}\Big)=0\nonumber
\end{eqnarray}
if $\nu_m \tau$ is a multiple of $2\pi$, and for $n$ even.
This means that a qubit can be left in a product state with a specific motional mode $m$ regardless of the values of $\tau_a$ and $\tau_b$. Unfortunately, the two motional modes in our system have incommensurable oscillation frequencies (note that $\nu_2/\nu_1=\sqrt{3}$) which leads to the impossibility of finding a $\tau$ that, independetly of $\tau_a$ and $\tau_b$,  decouples the qubits from both vibrational modes.

An AXY-4 sequence of a duration $4\tau$ such that $\tau=2\pi r/\nu_1$, makes $G_{j1}(4\tau)=0$ for any choice of $\tau_a$ and $\tau_b$, while we will numerically look for the values of $\tau_a$ and $\tau_b$ that minimise $G_{j2}(4\tau)$. For the sake of simplicity in the presentation of this part, we consider $f_1(t)=f_2(t)$, i.e. the same sequence is simultaneously applied to both qubits leading to $G_{1m}=G_{2m}$. However,  when considering real pulses, we will not use simultaneous driving in order to efficiently eliminate crosstalk effects which leads to an optimal performance of the method, see Sec.~\ref{SecSeq}. In Fig.~\ref{Gplot}{\bf(a)} we give a contour color plot of $G_{j2}(4\tau)$ with $\tau=2\pi/\nu_1$  for all combinations of $\tau_a$ and $\tau_b$. The dark blue regions represent the  values of $\tau_a$ and $\tau_b$ that minimise the $G_{j2}(4\tau)$ functions. Then any pair of $\tau_{a,b}$ in that region defines a valid sequence for a two-qubit phase gate. At Fig.~\ref{Gplot}{\bf(b)}, we give the corresponding value for $\tilde{\varphi}(4\tau)$ of the resulting two-qubit gate (red panels). In Figs.~\ref{Gplot}{\bf(c)}, \ref{Gplot}{\bf(d)} and \ref{Gplot}{\bf(e)}, \ref{Gplot}{\bf(f)} the same procedure is shown for  $\tau=2\times2\pi/\nu_1$ and $\tau=3\times2\pi/\nu_1$, respectively, i.e. for values $r=2$ and $r=3$, obtaining several combinations of $\tau_a$ and $\tau_b$ that result in a phase gate. Finally, to recover the real phase $\varphi(t)$, we multiply $\tilde{\varphi}(4\tau)$ by $(\Delta_1/ \nu)^2=\frac{\hbar\gamma_e^2g_B^2}{64 M \nu^3}$, according to Eq.~(\ref{phase}), showing the dependance of $\varphi$ on $\nu$ and $g_B$.

\section{Taylored Sequences and Results}\label{SecSeq}

We will benchmark the performance of our microwave pulsed scheme by means of detailed numerical simulations. The total Hamiltonian governing the dynamics is $H+H_c$. In a rotating frame with respect to $H_0$ and after neglecting terms that rotate at a speed of tens of GHz (see appendix A and C for more details), the effective Hamiltonian reads
\begin{eqnarray}\label{simstart}
\nonumber H^{\rm I}(t)&=&  \Delta_1 (b e^{-i\nu_1 t}+b^\dag e^{+i\nu_1 t}) \sigma_1^z - \Delta_2(c e^{-i\nu_2 t}+c^\dag e^{+i\nu_2 t})\sigma_1^z\\
\nonumber &+&    \Delta_1 (b e^{-i\nu_1 t}+b^\dag e^{+i\nu_1 t}) \sigma_2^z - \Delta_2(c e^{-i\nu_2 t}+c^\dag e^{+i\nu_2 t})\sigma_2^z\\
\nonumber &+& \frac{\Omega_1(t)}{2} \sigma_1^{\phi_1} + \frac{\Omega_1(t)}{2} (\sigma_2^+ e^{i\delta_2 t} e^{i\phi_1} + {\rm H.c}.)\\
                  &+& \frac{\Omega_2(t)}{2} \sigma_2^{\phi_2} + \frac{\Omega_2(t)}{2} (\sigma_1^+ e^{i\delta_1 t} e^{i\phi_2} + {\rm H.c}.).
\end{eqnarray}
Here, $\sigma_j^{\phi_j} = \sigma_j^+ e^{i\phi_j}  +  \sigma_j^- e^{-i\phi_j}$, and the last two lines contain both the resonant terms giving rise to the $\pi$-pulses, i.e. $ \frac{\Omega_1(t)}{2} \sigma_1^{\phi_1}$ and $ \frac{\Omega_2(t)}{2} \sigma_2^{\phi_2}$, as well as crosstalk contributions of each $\pi$-pulse on the off-resonant ion. The latter are  $\frac{\Omega_1(t)}{2} (\sigma_2^+ e^{i\delta_2t} e^{i\phi_1} + {\rm H.c}.)$ and $\frac{\Omega_2(t)}{2} (\sigma_1^+ e^{i\delta_1t} e^{i\phi_2} + {\rm H.c}.)$, where $\delta_2 = - \delta_1 = \omega_2 - \omega_1$. We use Eq.~(\ref{simstart}) as the starting point of our simulations without any further assumptions. In addition, our numerical simulations include motional decoherence described by a Lindblad equation describing an enviroment at a temperature of 50 K as well as errors on $\Omega_{1,2}$, $\omega_{1,2}$, and $\nu$.

\begin{figure}[b]
\centering
\includegraphics[width=1\linewidth]{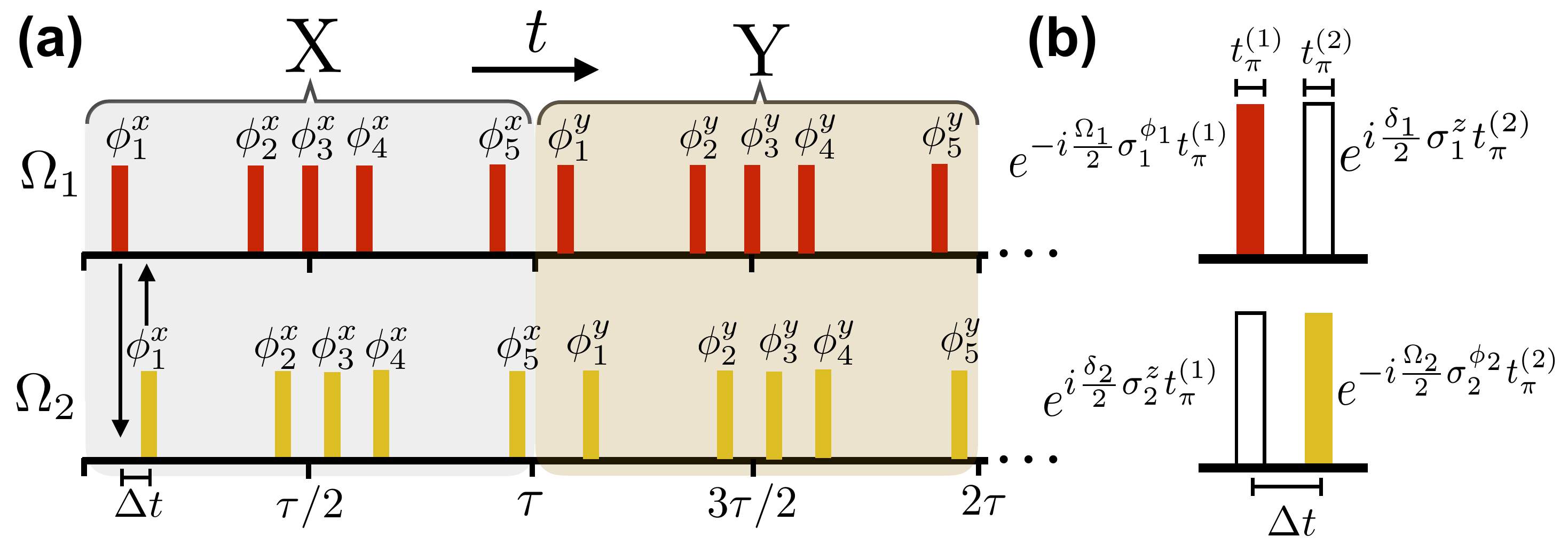}
\caption{{\bf (a)} Pulse sequence on the first(second) ion, upper(bottom) panel. The first(second) ion is driven with an AXY-4 sequence, red(yellow) blocks represent  $\pi$-pulses.  Each pulse on the second ion is separated by $\Delta t$ from the pulses acting on the first ion.  {\bf (b)} Zoom on two pulses. We can observe the propagators leading to $\pi$-pulses, i.e.  $e^{-i \frac{\Omega_{1,2}}{2} \sigma_{1,2}^{\phi_{1,2}} t_\pi }$
(red and yellow blocks), and their unwanted side-effects in the adjacent ions $e^{i\frac{\delta_{2,1}}{2} \sigma_{2,1}^{z} t_\pi}$ (empty blocks). }
\label{combinedAXY}
\end{figure}

To get rid of crosstalk effects, we use a decoupling scheme acting non simultaneously on both ions that, at the same time, meets conditions in Eqs.~(\ref{conditions}), and give rise to a tunable phase gate between the ions. In this respect, one can demonstrate that a term like 
\begin{equation}
 \frac{\Omega_1(t)}{2} \sigma_1^{\phi_1} + \frac{\Omega_1(t)}{2} (\sigma_2^+ e^{i\delta_2 t} e^{i\phi_1} + {\rm H.c}.),
\end{equation}
for a final time $t^{(1)}_\pi = \frac{\pi}{\Omega_1}$, i.e. the required time for a $\pi$-pulse on the first ion, has the associate propagator 
\begin{equation}
\label{crosstalk}
U_{t^{(1)}_\pi}=  e^{-i \frac{\Omega_1}{2} \sigma_1^{\phi_1} t_\pi } e^{i\frac{\delta_2}{2} \sigma_2^{z} t_\pi},
\end{equation}
if and only if the Rabi frequency $\Omega_1$ satisfies 
\begin{equation}
\Omega_1=\frac{\delta_2}{\sqrt{4k^2-1}}, \mbox{with} \ k\in \mathbb{N}.
\end{equation}
See appendix~F for a demonstration of this. In the same manner, the term $ \frac{\Omega_2(t)}{2} \sigma_2^{\phi_2} + \frac{\Omega_2(t)}{2} (\sigma_1^+ e^{i\delta_1 t} e^{i\phi_2} + {\rm H.c}.)$ 
gives rise to $U_{t^{(2)}_\pi} =  e^{-i \frac{\Omega_2}{2} \sigma_2^{\phi_2} t_\pi } e^{i\frac{\delta_1}{2} \sigma_1^{z} t_\pi}$ under the  conditions  $t^{(2)}_\pi = \frac{\pi}{\Omega_2}$ and 
$\Omega_2=|\delta_1|/\sqrt{4{k}^2-1}, \mbox{with} \ k\in \mathbb{N} $. Hence, when the microwave driving is applied non-simultaneously over the registers, one can clearly argue that a $\pi$-pulse on the first ion induces a dephasing-like propagator on the second ion (i.e. $e^{i\frac{\delta_2}{2} \sigma_2^{z} t_\pi}$) and vice versa. It turns out that the our dynamical decoupling sequence successfully eliminates such undesired contribution.

Two blocks of our non simultaneous AXY-n sequence is depicted in Fig.~\ref{combinedAXY}{\bf(a)}, where one has to select $\tau$, $\tau_{a,b}$ and $\Delta t$. While $\tau$, $\tau_{a,b}$, define the sequence acting on the first ion, a temporal translation $\Delta t$ of each $\pi$-pulse sets the sequence on the second ion. Note that $\Delta t$ must satisfy $\Delta t>t_{\pi}$, see Fig.~\ref{combinedAXY}{\bf(b)},  to assure there is no pulse overlap. As we said before, the construction in Fig.~\ref{combinedAXY}{\bf(a)} eliminates the dephasing terms $e^{i\frac{\delta_{2,1}}{2} \sigma_{2,1}^{z} t_\pi}$. For example, the propagator for the first ion after a XY block $U^{(1)}_{\rm XY}$,  upper panel in Fig.~\ref{combinedAXY}{\bf (a)}, reads
\begin{eqnarray}\label{firstprop}
U^{(1)}_{\rm XY}&=&\bigg[ e^{i\frac{\delta_1}{2}  \sigma_1^{z} t_\pi}\sigma_1^{\phi_5^y}\bigg]\bigg[e^{i\frac{\delta_1}{2} \sigma_1^{z} t_\pi}\sigma_1^{\phi_4^y}\bigg]\bigg[e^{i\frac{\delta_1}{2} \sigma_1^{z} t_\pi}\sigma_1^{\phi_3^y}\bigg] \bigg[ e^{i\frac{\delta_1}{2} \sigma_1^{z} t_\pi}\sigma_1^{\phi_2^y}\bigg]\nonumber\\
&&\bigg[ e^{i\frac{\delta_1}{2} \sigma_1^{z} t_\pi}\sigma_1^{\phi_1^y}\bigg]\bigg[e^{i\frac{\delta_1}{2} \sigma_1^{z} t_\pi}\sigma_1^{\phi_5^x}\bigg] \bigg[ e^{i\frac{\delta_1}{2} \sigma_1^{z} t_\pi}\sigma_1^{\phi_4^x}\bigg] \bigg[e^{i\frac{\delta_1}{2} \sigma_1^{z} t_\pi} \sigma_1^{\phi_3^x} \bigg]\nonumber \\
&&\bigg[e^{i\frac{\delta_1}{2} \sigma_1^{z} t_\pi} \sigma_1^{\phi_2^x} \bigg]\bigg[e^{i\frac{\delta_1}{2} \sigma_1^{z} t_\pi} \sigma_1^{\phi_1^x}\bigg]\nonumber \\
&=& \sigma_1^{\phi_5^y} \sigma_1^{\phi_4^y} \sigma_1^{\phi_3^y} \sigma_1^{\phi_2^y} \sigma_1^{\phi_1^y} \sigma_1^{\phi_5^x} \sigma_1^{\phi_4^x} \sigma_1^{\phi_3^x} \sigma_1^{\phi_2^x} \sigma_1^{\phi_1^x}.
\end{eqnarray}
where the last equality is achieved using $\{\sigma_1^z,   \sigma_1^{\phi_j^{x,y}}\} = 0$.  Equation~(\ref{firstprop}) describes a situation without motional degrees of freedom. However the cancelation of the dephasing terms is still valid if one includes the spin-motion coupling terms because they depend on $\sigma^z_{1,2}$, see Eq.~(\ref{Hamiltonianbare}), and the operators $e^{i\frac{\delta_{1,2}}{2} \sigma_{1,2}^z t_{\pi}}$ commutes with them leading to the same cancelation.

In the same manner, one can find the propagator for the second ion $U^{(2)}_{\rm XY}$, see lower panel in  Fig.~\ref{combinedAXY}{\bf (a)}. This propagator reads
\begin{eqnarray}\label{secondprop}
U^{(2)}_{\rm XY}&=&\bigg[\sigma_2^{\phi_5^y} e^{i\frac{\delta_2}{2}  \sigma_2^{z} t_\pi}\bigg]\bigg[\sigma_2^{\phi_4^y} e^{i\frac{\delta_2}{2} \sigma_2^{z} t_\pi}\bigg]\bigg[\sigma_2^{\phi_3^y} e^{i\frac{\delta_2}{2} \sigma_2^{z} t_\pi}\bigg] \bigg[\sigma_2^{\phi_2^y} e^{i\frac{\delta_2}{2} \sigma_2^{z} t_\pi}\bigg]\nonumber\\
&&\bigg[ \sigma_2^{\phi_1^y} e^{i\frac{\delta_2}{2} \sigma_2^{z} t_\pi}\bigg]\bigg[\sigma_2^{\phi_5^x} e^{i\frac{\delta_2}{2} \sigma_2^{z} t_\pi}\bigg] \bigg[\sigma_2^{\phi_4^x} e^{i\frac{\delta_2}{2} \sigma_2^{z} t_\pi}\bigg] \bigg[\sigma_2^{\phi_3^x} e^{i\frac{\delta_2}{2} \sigma_2^{z} t_\pi}  \bigg]\nonumber \\
&&\bigg[\sigma_2^{\phi_2^x} e^{i\frac{\delta_2}{2} \sigma_2^{z} t_\pi}\bigg]\bigg[ \sigma_2^{\phi_1^x}e^{i\frac{\delta_2}{2} \sigma_2^{z} t_\pi}\bigg]\nonumber \\
&=& \sigma_2^{\phi_5^y} \sigma_2^{\phi_4^y} \sigma_2^{\phi_3^y} \sigma_2^{\phi_2^y} \sigma_2^{\phi_1^y} \sigma_2^{\phi_5^x} \sigma_2^{\phi_4^x} \sigma_2^{\phi_3^x} \sigma_2^{\phi_2^x} \sigma_2^{\phi_1^x}.
\end{eqnarray}

We can see that after an XY block, there is no contribution of dephasing like operators, see the last lines in Eqs.~(\ref{firstprop}) and~(\ref{secondprop}). Hence, a sequence XYXY applied to both ions following the scheme in Fig.~\ref{combinedAXY}{\bf (a)} will also share this property with the additional advantage of being robust against control errors~\cite{Casanova15}.

After simulating the application of a non-simultaneous AXY-4 sequence, we show the results (infidelities) in Table~\ref{table1}. It is noteworthy to comment that our numerical results have been calculated including motional decoherence. More specifically, we have added to the dynamics governed by the Hamiltonian in Eq.~(\ref{simstart})  a dissipative term of the form, see for example~\cite{Brownnutt15},
\begin{eqnarray}
D(\rho) &=& \frac{\Gamma_b}{2}(\bar{N}_b+1) (2 b \rho b^{\dag} - b^{\dag} b \rho - \rho b^{\dag} b) \nonumber\\
 &+& \frac{\Gamma_b}{2} \bar{N}_b (2 b^\dag \rho b - b  b^{\dag} \rho - \rho b b^{\dag})\nonumber\\
&+&\frac{\Gamma_c}{2}(\bar{N}_c+1) (2 c \rho c^{\dag} - c^{\dag} c \rho - \rho c^{\dag} c) \nonumber\\
&+& \frac{\Gamma_c}{2} \bar{N}_c (2 c^\dag \rho c - c  c^{\dag} \rho - \rho c c^{\dag}), 
\end{eqnarray}
where an estimation of the values for the heating rates $\Gamma_{b,c}$ is given in appendix~G for each of the specific examples considered in the main text, while $\bar{N}_{b,c} \equiv N^{\rm thermal}_{b,c} =1/(e^{\hbar \nu_{1,2}/k_{\rm B} T}-1)$ where we  have considered a temperature of $T=50$K.

\begin{table}[t!]
\centering
\caption{Infidelities ($I$) for two-qubit gates after the application of 20 imperfect microwave pulses on each ion, according to our AXY-4 protocol, for several initial states, $\psi_j$, and different experimental conditions, see main text. We focus in $\pi/4$ and $\pi/8$ entangling phase gates, however our method is general and can achieve any phase. Initial states, up to normalization, are $\psi_{1}=|{\rm g}\rangle \otimes (|{\rm g}\rangle + |{\rm e}\rangle)$, $\psi_{2}= (|{\rm g}\rangle + |{\rm e}\rangle) \otimes (|{\rm g}\rangle + |{\rm e}\rangle)$, $\psi_{3}=|{\rm g}\rangle \otimes (|{\rm g}\rangle + i |{\rm e}\rangle) + |{\rm e}\rangle \otimes |{\rm e}\rangle$, $\psi_{4}=|{\rm e}\rangle \otimes (|{\rm g}\rangle - i |{\rm e}\rangle) + |{\rm g}\rangle \otimes |{\rm g}\rangle$, and $\psi_{5}=|{\rm e}\rangle \otimes (|{\rm g}\rangle - i |{\rm e}\rangle) + |{\rm g}\rangle \otimes (|{\rm g}\rangle + i |{\rm e}\rangle)$.}
\label{table1}
\vspace{2.0mm}
\begin{tabular}{{ |c | c | c | c | c| c|}}
\hline
$I$ ($\times 10^{-4}$) &exp$( i\frac{\pi}{4} \sigma_1^z  \sigma_2^z)$ &exp$( i\frac{\pi}{8} \sigma_1^z  \sigma_2^z)$&exp$( i\frac{\pi}{4} \sigma_1^z  \sigma_2^z)$ &exp$( i\frac{\pi}{8} \sigma_1^z  \sigma_2^z)$\\
 \hline
$\psi_1$  & $1.172$    &$0.128$      & $2.060$   &$0.144$  \\
\hline
$\psi_2$     & $2.229 $    &$0.136 $      & $4.905$   & $0.304 $ \\
\hline
$\psi_3$  & $3.052$    &$0.116  $    & $5.899 $  & $0.371$ \\
\hline
$\psi_4$  & $ 4.631 $    &$0.172  $    & $5.946$  & $0.413$ \\
\hline
$\psi_5$  & $3.250$    &$0.110 $    & $4.635 $  & $0.293$ \\
\hline
\end{tabular}
\end{table}

We computed the gate infidelity for the following situations. Firstly, we simulated the gates exp$( i\frac{\pi}{4} \sigma_1^z  \sigma_2^z)$ and exp$( i\frac{\pi}{8} \sigma_1^z  \sigma_2^z)$, second and third columns in Table~\ref{table1}, in a time of $80 \ \mu$s for a magnetic field gradient of $g_B=150 {\rm \frac{T}{m}}$~\cite{Weidt16}.  We designed the microwave sequence such that $\tau= 3\times2\pi r/\nu_1$ leading to a gate time which is 12 times the period of the com mode.  Other relevant parameters are $\nu_1=\nu_2/\sqrt{3} = (2\pi)\times 150$ kHz, $\pi$-pulse time of $\approx 75$ ns that implies a Rabi frequency of $\Omega_1=\Omega_2=\Omega\approx (2\pi)\times 6.63$ MHz, and $\omega_2 - \omega_1 = (2\pi)\times 25.7$ MHz, while we have chosen $\Delta t$ as 1.05 times the $\pi$-pulse time. The bosonic modes, $b$ and $c$, are initially in a thermal state with $0.2$ phonons each~\cite{Weidt15}. In addition to heating processes with rates $\Gamma_b \bar{N}_b \approx (2\pi) \times 133$ Hz and $\Gamma_c \bar{N}_c \approx (2\pi) \times 9$ Hz (appendix~G), our simulations include a Rabi frequency mismatch of $1\%$, a trap frequency shift of $0.1\%$, and an energy shift of $(2\pi)\times 20$ kHz on both ions.

Secondly, we also target the gates exp$( i\frac{\pi}{4} \sigma_1^z  \sigma_2^z)$ and exp$( i\frac{\pi}{8} \sigma_1^z  \sigma_2^z)$, fourth and fifth columns in Table~\ref{table1}, but now with $g_B=300 {\rm \frac{T}{m}}$. The gate time is $36.3 \ \mu$s, i.e. 8 times the oscillation period of the com mode whose frequency is $\nu=\nu_1=\nu_2/\sqrt{3} = (2\pi)\times 220$ kHz. Other parameters are $\Omega\approx (2\pi)\times 10$ MHz,  $\pi$-pulse time of $\approx 49$ ns, $\omega_2 - \omega_1 = (2\pi)\times 39.8$ MHz and the energy shift upon the ions, errors on Rabi and trap frequencies,  $\Delta t$, and the initial bosonic states are the same as in the previous case. Because of the new value for $g_B$, the
heating rates had to be recalculated leading to $\Gamma_b \bar{N}_b \approx (2\pi) \times 248$ Hz  and $\Gamma_c \bar{N}_c \approx (2\pi) \times16$ Hz.

In Table~\ref{table1} we find that, even in the presence of the errors we have included, our method leads to fast two-qubit gates with fidelities exceeding 0.999. Finally, we note that higher values of $g_B$ will result in faster gates.

\section{conclusions}

We have demonstrated that pulsed DD schemes are efficient generators of fast and robust two-qubit gates. Our microwave sequence  forces the two motional modes in a certain direction to cooperate, and makes the gate fast and robust against external noise sources including motional heating. This novel technique opens a path in the microwave control of trapped ions, and can be generalised to laser-based setups.

\section*{acknowledgments}
This work was supported by the ERC Synergy grant BIOQ (grant no. 319130), the EU
Projects EQUAM and DIADEMS, the SFB/TRR 21, Spanish MINECO/FEDER FIS2015-69983-P, Basque Government IT986-16, and PhD grant PRE-2015-1-0394. J.~C. acknowledges Universität Ulm for a Forschungsbonus.

\section*{Appendix A: Initial approximations}

\subsubsection{Two-Level Approximation}
In this section we numerically argue that the presence of the additional hyperfine levels of the $^{171}$Yb$^+$ ion, the fluctuations of the magnetic field, and the effect of fast rotating terms does not threaten the gate fidelities claimed in this article.  For numerical simplicity we have considered a single four level system and looked for the fidelity of the propagator after a  sequence of 20 $\pi$-pulses is applied. We find that the error (infidelity) is on the order of $10^{-5}$, hence being one order of magnitude below the gate errors reported in Table I of the main text. Therefore, we conclude that the presence of the additional levels, counter rotating terms, and the fluctuations of the magnetic field have a negligible effect on the final fidelity of the gate to the order claimed in the main text. We detail now the parameters and conditions in our numerical simulations. 

In the hyperfine ground state of the $^{171}$Yb$^+$ ion, transitions can be selected with the appropriate polarization of the control fields. However, experimental imperfections might generate unwanted leakage of population from the qubit-states to other states. On the other hand, the presence of fluctuations of the magnetic field may also result in imperfect $\pi$-pulses which may also damage the performance of the gate. To account for these experimental imperfections we simulate the following 4-level Hamiltonian
\begin{eqnarray}
H_{4l}&=&E_0 |0\rangle \langle 0 | + E_1|1\rangle \langle 1 | + E_2 |2\rangle \langle 2 | + E_3 |3\rangle \langle 3 |  \\
&+& X(t) |1\rangle \langle 1 | -X(t) |3\rangle \langle 3 | \nonumber \\
&+& \Omega(t) ( | 0 \rangle \langle 1| + \epsilon | 0 \rangle \langle 2 | + \epsilon | 0 \rangle \langle 3 | +{\rm H. c.} ) \cos{[\omega t + \phi(t)]}, \nonumber 
\end{eqnarray}
where the energies of the hyperfine levels, $E_i$, are those corresponding to a $^{171}$Yb$^+$ ion in a magnetic field of $100$ G, and the qubit is codified in levels $ \{|0 \rangle, | 1 \rangle \}$. Function $X(t)$ represents a fluctuating magnetic field, which shifts the magnetically sensitive levels $|1\rangle $ and $|3\rangle$ in opposite directions. Numerically we have constructed this function as an Ornstein-Uhlenbeck (OU) process~\cite{Gillespie96}, where the parameters have been chosen such that the qubit-levels, in the absence of any pulses, show a coherence decaying exponentially with a $T_2$ coherence-time of $3$ms, as experimentally observed~\cite{Wineland98}. Particularly, this corresponds to values  $\tau=50 \  {\rm \mu s}$ for the correlation time, and $c=2/(\tau*T_2)$ for the diffusion constant of the OU process. $\Omega(t)$ is a step function taking exclusively values $\Omega$ and $0$, and $\epsilon$ is a small number representing the leaking of the qubit population through unwanted transitions. For the numerical analysis we have used unfavourable values for this set of parameters. More specifically, the Rabi frequency was assigned a value of $\Omega=(2\pi)\times20$MHz, which is already twice the maximum value used in all the other simulations throughout the article, having therefore a larger probability of exciting other, undesired, hyperfine transitions. Moreover, the simulations were performed for the longest sequence discussed in the main text, which lasts $80\  {\rm \mu s}$. 

\begin{figure}[t] 
\centering
\includegraphics[width=1\linewidth]{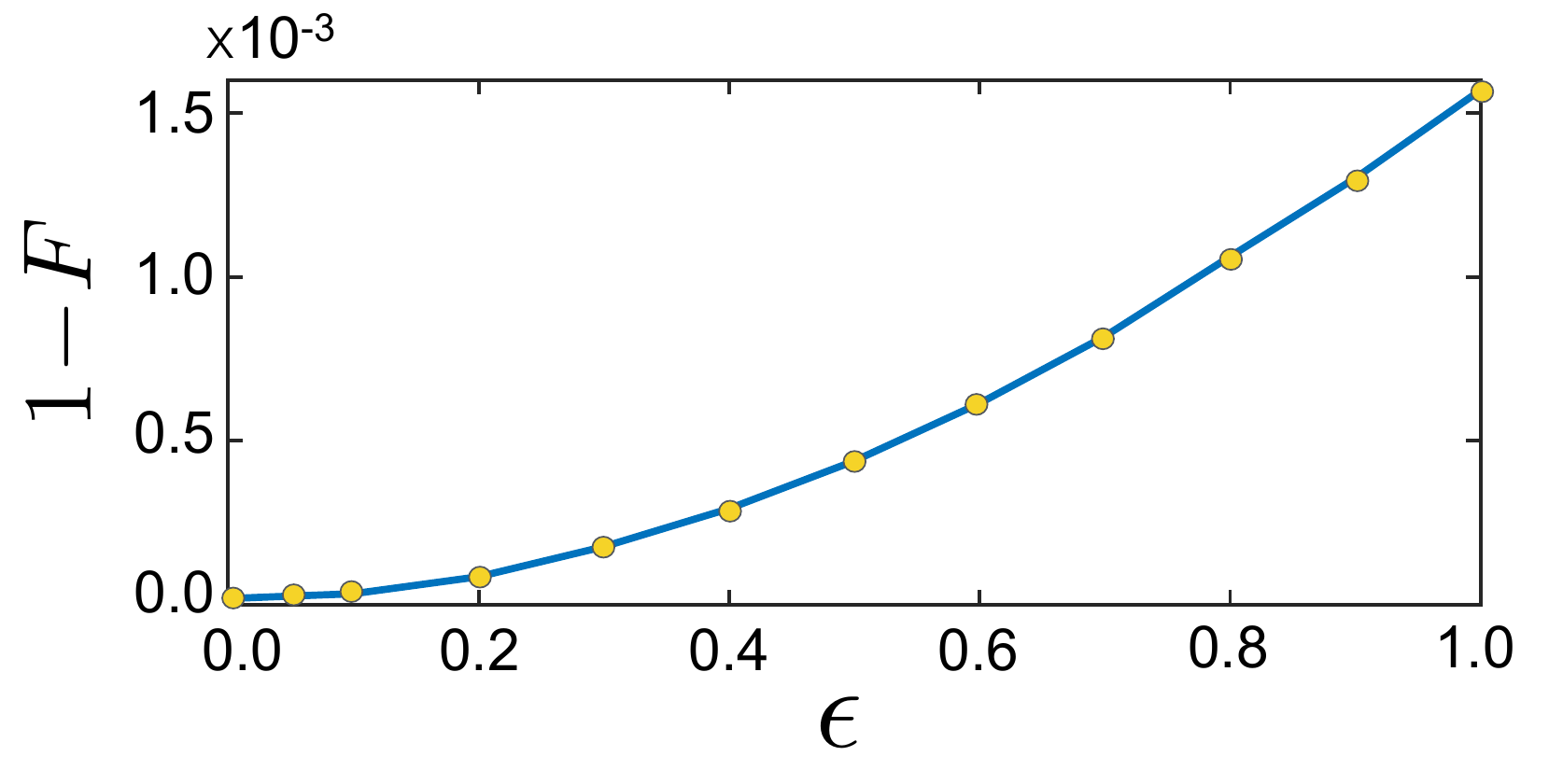}
\caption{Infidelity of an AXY-4 sequence, consisting of 20 $\pi$-pulses and a total time of $80 {\rm \mu s}$, vs the strength of the leakage of population to other spectator levels. Each point is the average of the infidelities of 100 runs of the sequence in the presence of stochastic fluctuations of the magnetic field. $\epsilon$ represent the strength of the transitions to unwanted levels in the hyperfine manyfold of $^{171}$Yb$^+$, expressed as a fraction of the Rabi frequency $\Omega$}
\label{Infidelity}
\end{figure}
We compare the propagator resulting from our simulations to the identity, which is what one would expect after an even number of $\pi$-pulses, 20 in our case, and we compute a value for the fidelity according to the definition
\begin{equation}
F_{A,B}=\frac{|{\rm Tr}(AB^\dag)|}{\sqrt{{\rm Tr}(AA^{\dag}) {\rm Tr} (BB^\dag)}},
\end{equation}
where $F_{A,B}$ is the fidelity between operators $A$ and $B$. To account for the stochastic effects of the OU process that models the fluctuations of the magnetic field, we have averaged the resulting fidelities over $100$ runs of our numerical simulator. In Fig.~\ref{Infidelity} we show the value of the infidelity, $1-F$, for a number of values of $\epsilon$. We can see that the error grows non-linearly with the strenght of the leakage due to polarisation errors of the control fields. However, for alignment errors below $20\%$ ($\epsilon=0.2$) we obtain that the infidelity is smaller than $10^{-4}$. Hence, for polarisation errors below $20\%$, the effect of additional hyperfine levels, magnetic field fluctuations, and fast counter rotating terms should only be detectable in the fifth significant order of the gate fidelity, and not alter the $99.9\%$ fidelity claimed in the article.

\subsubsection{Coupling with radial modes}

\begin{figure}[t]
\centering
\includegraphics[width=1\linewidth]{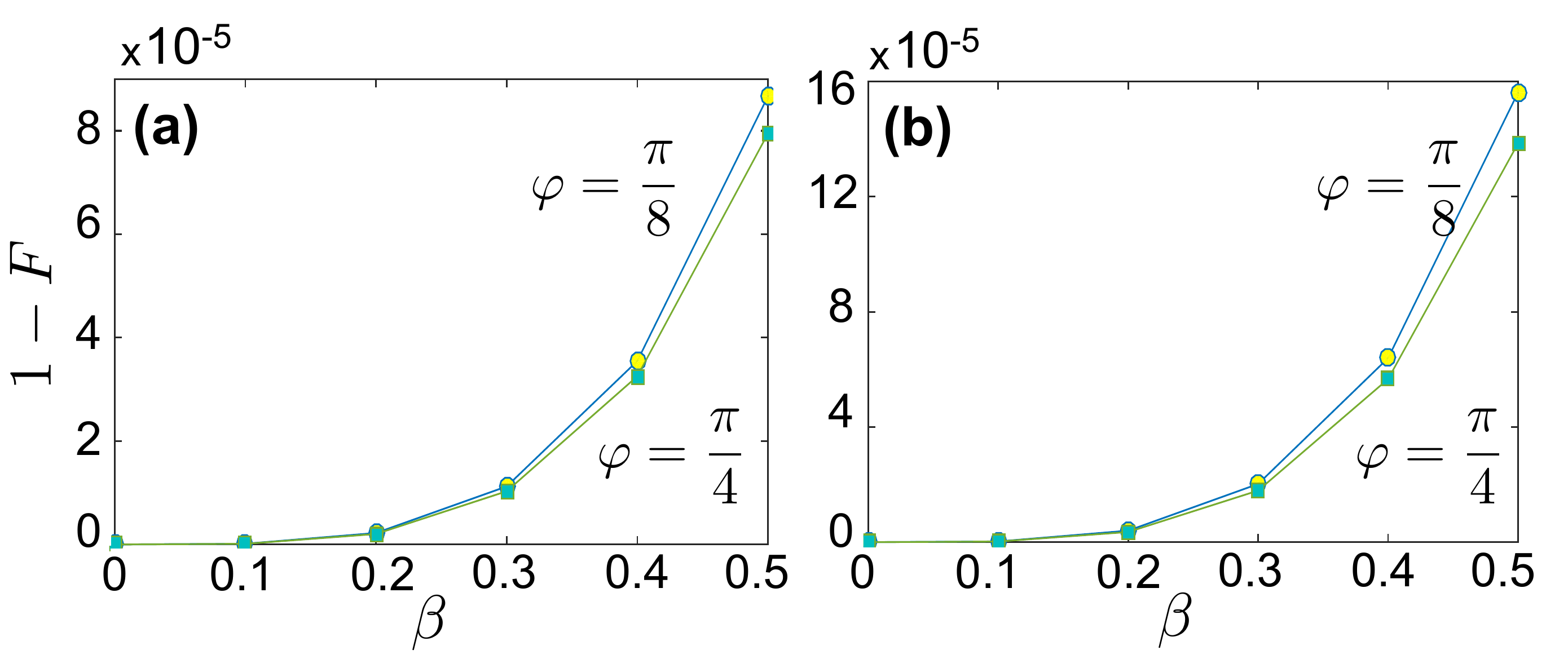}
\caption{Final state infidelity as a consequence of the presence of coupling with a radial mode of the kind in Eq.~(\ref{radial}). We observe a growing infidelity for larger values of $\beta$ with $\beta=\Delta_r/\Delta_1$. {\bf(a)} The case of $g_B=150 {\rm \frac{T}{m}}$ for two qubit gate phases $\varphi=\pi/4$ squares, and $\varphi=\pi/8$ circles. In {\bf (b)} we use $g_B= 300 {\rm \frac{T}{m}}$ and, again, we use squares for $\varphi=\pi/4$, and circles for $\varphi=\pi/8$. For both plots we used $\psi_4$, see the main text, as the initial state.}
\label{radialinfidelity}
\end{figure}

In this section we study the influence of the motional radial modes of the ion in our proposal. To account for the effect of a given radial mode $d$, Hamiltonian~(1) of the main text needs to be complemented with a term of the form
\begin{eqnarray}\label{radial}
 \nu_r d^{\dag}d + \Delta_r (d  + d^\dag) [ \sigma_1^z  +  \sigma_2^z].
\end{eqnarray}
Because of computational restrictions, in this analysis we will only consider one radial mode, and assume no motional decoherence. Term~(\ref{radial}) is justified because of the unavoidable presence of some remanent magnetic field gradient in the radial direction, which leads to the coupling $\Delta_r$ that we model as a fraction of the $\Delta_1$ coupling in Hamiltonian~(1) of the main text, i.e. $\Delta_r= \beta\Delta_1$. We have compared the states evolved under Hamiltonian~(1) in the main text including and not the coupling term in Eq.~(\ref{radial}), and computed the infidelity between them. In Fig.~\ref{radialinfidelity}  we show the results for different values of $\beta$. The value of $\nu_r = 2.5$ MHz was used in the simulations, and the initial state for the qubits was chosen to be $\psi_4$ in Table I, while a thermal state with 2 phonons was used as the initial state of the radial mode. We observe that even for large values of $\beta$, the impact of the radial mode is negligible, on the order of $10^{-5}$ for values of $\beta$ up to 0.4, which are experimentally unexpected.

\section*{Appendix B: two hyperfine ions under a magnetic field gradient}

The Hamiltonian of the relevant hyperfine levels of the two qubit system (composed, in our case, of two $^{171}$Yb$^+$ ions) under a $z$ dependent magnetic field can be expressed as $(\hbar = 1)$
\begin{eqnarray}\label{model}
H &=&     [\omega_{\rm{e}} + \gamma B(z_1)] |{\rm{e}}\rangle \langle {\rm{e}}|_1 +  \omega_{\rm{g}} |{\rm{g}} \rangle \langle {\rm{g}}|_1  \nonumber\\
      &+&   [\omega_{\rm{e}} + \gamma B(z_2)] |{\rm{e}}\rangle \langle {\rm{e}}|_2 +  \omega_{\rm{g}} |{\rm{g}} \rangle \langle {\rm{g}}|_2 \nonumber\\ 
      &+& \nu_1 a^\dag a + \nu_2 c^\dag c, 
\end{eqnarray}
where $\gamma$ relates to the electronic gyromagnetic ratio as $\gamma \equiv \frac{|\gamma_e|}{2} \approx (2\pi) \times 1.4$ MHz/G, and $B(z_j)$ is a position-dependent magnetic field that generates an additional energy splitting on the first $(j =1)$ and second $(j=2)$ ions. We have assumed that the ions, which interact through direct Coulomb force, perform only small oscillations around their equilibrium positions, $z_j=z_j^0+q_j$. Under this assumption, Hamiltonian diagonalization yields frequencies $\nu_1 = \nu$ and $\nu_2 = \sqrt{3} \nu$, $\nu$ being the axial trapping frequency, for the collective normal modes, namely the center-of-mass and breathing modes~\cite{James98}. 

If $B(z_{j})$ can be expanded to a good approximation to the first order in $q_j$, then $B(z_j)=B_{j} + g_{B}  \ q_{j}$, where $B_j\equiv B(z_j^0)$ and $g_B\equiv \partial B/{\partial z_j}\big|_{z_j=z_j^0}$. With this, and up to an energy displacement, the Hamiltonian~(\ref{model}) reads 
\begin{eqnarray}\label{Hamil2}
H &=& \frac{1}{2}[\omega_{\rm{e}} + \gamma B_1 - \omega_{\rm g}] \ \sigma_1^z + \frac{1}{2}[\omega_{\rm{e}} + \gamma B_2 - \omega_{\rm g}] \ \sigma_2^z \\
&+& \nu_1 a^{\dag}a +  \nu_2 c^{\dag}c +\frac{\gamma g_{B}}{2}(q_1+q_2) +\frac{\gamma g_{B}}{2}(q_1\sigma_1^z+q_2\sigma_2^z), \nonumber
\end{eqnarray}
where we have used the relations $|{\rm{e}}\rangle \langle {\rm{e}}|_j=\frac{1}{2}(\mathds{1}+\sigma_{\! j} ^z)$ and $|{\rm{g}}\rangle \langle {\rm{g}}|_j=\frac{1}{2}(\mathds{1}-\sigma_{\! j}^z)$. At this moment, it may be useful to recall that the displacement of the ions from their equilibrium positions, $q_1$ and $q_2$ can be expressed in terms of the collective normal modes, $Q_1$ and $Q_2$, as
\begin{eqnarray}\label{quantizedpositions}
q_1&=& \frac{Q_1 -  Q_2}{\sqrt{2}}= \sqrt{\frac{\hbar}{4M\nu_1}} \big(a+a^\dag\big) - \sqrt{\frac{\hbar}{4M\nu_2}} \big(c+c^\dag\big), \nonumber \\ 
q_2&=& \frac{Q_1 + Q_2}{\sqrt{2}}= \sqrt{\frac{\hbar}{4M\nu_1}} \big(a+a^\dag\big) + \sqrt{\frac{\hbar}{4M\nu_2}} \big(c+c^\dag\big),
\end{eqnarray}
$M$ being the mass of each ion. Using these relations, which follow the prescription in~\cite{James98}, Eq.(\ref{Hamil2}) can be rewritten as
\begin{eqnarray}\label{Hamil3}
H &=& \frac{1}{2}[\omega_{\rm{e}} + \gamma B_1 - \omega_{\rm g}] \ \sigma_1^z + \Delta_1(a + a^\dag) \ \sigma_1^z - \Delta_2 (c + c^\dag)\ \sigma_1^z\nonumber\\
&+& \frac{1}{2}[\omega_{\rm{e}} + \gamma B_2 - \omega_{\rm g}] \ \sigma_2^z + \Delta_1(a + a^\dag) \ \sigma_2^z + \Delta_2(c + c^\dag)\ \sigma_2^z\nonumber\\
&+& \nu_1 a^{\dag}a +  \nu_2 c^{\dag}c + \frac{\gamma g_{B}}{2} \sqrt{\frac{\hbar}{M \nu_1}} (a + a^\dag),
\end{eqnarray}
where we have defined $\Delta_{1,2} \equiv  \frac{\gamma g_{B}}{4}\sqrt{\frac{\hbar}{M \nu_{1,2}}}$ as the coupling strength between the qubits and the normal modes. The last term in Eq.(\ref{Hamil3}) can be absorbed by a redefined bosonic operator $b = a + \alpha$, with $\alpha = 2\Delta_1/\nu_1$, which results in Hamiltonian
\begin{eqnarray}\label{simplifiyedJorge}
 \nonumber H&=& \frac{\omega_1}{2}\sigma_1^z + \Delta_1 (b+b^\dag) \sigma_1^z - \Delta_2(c+c^\dag)\sigma_1^z\\
     &+& \frac{\omega_2}{2}\sigma_2^z + \Delta_1 (b+b^\dag) \sigma_2^z + \Delta_2(c+c^\dag)\sigma_2^z\\
    &+& \nu_1 b^\dag b + \nu_2 c^\dag c\nonumber,\end{eqnarray}
where $\omega_{1,2} \equiv \omega_{\rm{e}} - \omega_{\rm g}  - 2\alpha\Delta_1  + \gamma B_{1,2}$. Furthermore, we can easily compute the quantity $\omega_{2} - \omega_{1}=\gamma (B_{2} - B_{1})=\gamma g_B (z_2-z_1) = \gamma g_B \Delta z$.

\section*{ Appendix C: The Interaction Hamiltonian}

A bichromatic microwave field of frequencies $\omega_j$ and phase $\phi_j$ will be applied to the system described by Eq.(\ref{simplifiyedJorge}). The action of such microwave field on the ions is described by the following Hamiltonian
\begin{equation}\label{MWField}
H_c(t)=\sum_{j=1}^2\Omega_j(t) (\sigma_1^x + \sigma_2^x)\cos(\omega_j t - \phi_j)
\end{equation}
where $\Omega_j$ is the Rabi frequency associated to the intensity of the microwave field with frequency $\omega_j$. If we add this term to the Hamiltonian (\ref{simplifiyedJorge}), and we move to an interaction picture with respect to $H_0=\frac{\omega_1}{2} \sigma_1^z + \frac{\omega_2}{2} \sigma_2^z + \nu_1 b^{\dag}b +  \nu_2 c^{\dag}c$, the complete Hamiltonian in the interaction picture will be given by
\begin{widetext}
\begin{eqnarray}\label{MWField2}
H^{\rm I}(t)=e^{iH_0t}H_{int}(t)e^{-iH_0t} =
\Delta_1b e^{-i \nu_1 t} \sigma_1^z - \Delta_2ce^{-i\nu_2 t} \sigma_1^z + \Delta_1b e^{-i \nu_1 t} \sigma_2^z +\Delta_2ce^{-i\nu_2 t} \sigma_2^z \nonumber \\
+ \big[\Omega_1(t)\cos(\omega_1 t - \phi_1) +\Omega_2(t)\cos(\omega_2 t - \phi_2) \big](\sigma_1^+ e^{i \omega_1 t} + \sigma_2^+ e^{i \omega_2 t} ) +{\rm H.c.}
\end{eqnarray}
where we have use the relations $e^{i\theta a^\dagger a} a e^{-i\theta a^\dagger a}=ae^{-i\theta}$ and $e^{i\theta \sigma^z} \sigma^+ e^{-i\theta \sigma^z}=\sigma^+e^{i2\theta}$. Rewriting the last term leads to 
\begin{eqnarray}\label{MWField3}
H^{\rm I}(t)&=&\Delta_1b e^{-i \nu_1 t} \sigma_1^z - \Delta_2ce^{-i\nu_2 t} \sigma_1^z + \Delta_1b e^{-i \nu_1 t} \sigma_2^z +\Delta_2ce^{-i\nu_2 t} \sigma_2^z \nonumber \\
&+& \frac{\Omega_1(t)}{2}\Big[\sigma_1^+ e^{i \phi_1}(e^{i2(\omega_1 t -\phi_1)} + 1) + \sigma_2^+( e^{i(\omega_1+ \omega_2) t}e^{-i\phi_1} + e^{-i(\omega_1-\omega_2)t}e^{i\phi_1}) \Big] \nonumber\\
&+& \frac{\Omega_2(t)}{2}\Big[\sigma_1^+( e^{i(\omega_2+ \omega_1) t}e^{-i\phi_2} + e^{-i(\omega_2-\omega_1)t}e^{i\phi_2})  + \sigma_2^+  e^{i \phi_2}(e^{i2(\omega_2 t -\phi_2)} + 1)\Big] 
+{\rm H.c.}
\end{eqnarray}
\end{widetext}
At this point we can safely neglect the terms that rotate with frequencies $\pm|2\omega_1|,\pm|2\omega_2|$ and $\pm|\omega_1+\omega_2|$ by invoking the RWA. Because $|\omega_1|,|\omega_2| \gg \Omega_1,\Omega_2$, these terms will have a negligible effect on the evolution of the system. On the other hand, terms that rotate with frequencies $\pm|\omega_2-\omega_1|$ would have a significant effect on the evolution of the system. However this will be suppressed at the end of the two-qubit gate, because of the design of the pulse sequence. How this elimination occurs is covered in Sec.\ref{PulseP}. Hence, we can assume that these terms do not have any effect on the system and we can neglect them, thus the Hamiltonian is 
\begin{eqnarray}\label{MWField4}
 H^{\rm I}(t)&=& \Delta_1(b e^{-i \nu_1 t} + b^\dag e^{i\nu_1 t}) \sigma_1^z - \Delta_2(ce^{-i\nu_2 t} + c^\dag e^{i \nu_2 t}) \sigma_1^z \\ \nonumber &+& \Delta_1(b e^{-i \nu_1 t} + b^\dag e^{i\nu_1 t}) \sigma_2^z + \Delta_2(ce^{-i\nu_2 t} + c^\dag e^{i \nu_2 t}) \sigma_2^z\\
\nonumber &+& \frac{\Omega_1(t)}{2}(\sigma_1^+ e^{i \phi_1} + \sigma_1^- e^{-i\phi_1}) + \frac{\Omega_2(t)}{2}(\sigma_2^+ e^{i \phi_2} + \sigma_2^- e^{-i\phi_2}),
\end{eqnarray}
which corresponds to Eq.(2) in the main text.

\section*{Appendix D: The time evolution operator}

An analytical expression for the time evolution operator exits for any Hamiltonian of the form
\begin{eqnarray}\label{Hdd2}
H^{\rm II}(t) = \sum_{j=1}^N\sum_{m=1}^M f_{j}(t)\Delta_{jm} (a_m e^{-i\nu_mt} + a_m^\dag e^{i\nu_mt}) \ \sigma_j^z.
\end{eqnarray}
The Hamiltonian for our ion system, undergoing a sequence of instantaneous $\pi$-pulses, is given by
\begin{eqnarray}\label{Hdd}
H^{\rm II}(t) &=&  f_{1}(t)\Delta_1 (b e^{-i\nu_1t} + b^\dag e^{i\nu_1t}) \ \sigma_1^z \nonumber\\
&-& f_{1}(t)\Delta_2 (c e^{-i\nu_2t} + c^\dag e^{i\nu_2t})\ \sigma_1^z\nonumber\\   &+& f_{2}(t)\Delta_1 (b e^{-i\nu_1t} + b^\dag e^{i\nu_1t}) \ \sigma_2^z \nonumber\\
 &+& f_{2}(t)\Delta_2 (c e^{-i\nu_2t} + c^\dag e^{i\nu_2t})\ \sigma_2^z,
\end{eqnarray}
and therefore belongs to this category with $a_1=b$, $a_2=c$ and $\Delta_{11}=\Delta_{21}=\Delta_1$, $\Delta_{12}=-\Delta_{22}=-\Delta_2$.

The time evolution operator of a time dependent Hamiltonian is given by the Dyson series or equivalently by the Magnus expansion:
\begin{eqnarray}\label{Magnus}
U(t) = \exp{\big\{\Omega_1(t) +\Omega_2(t)+\Omega_3(t)+...\big\}},
\end{eqnarray}
where (in general for $t_0\neq0$)
\begin{eqnarray}\label{MagnusTerms}
\Omega_1(t,t_0) &=& -i\int_{t_0}^t dt_1H(t_1)\nonumber\\
\Omega_2(t,t_0) &=& -\frac{1}{2}\int_{t_0}^{t}dt_1\int_{t_0}^{t_1} dt_2 [H(t_1),H(t_2)] \\
\Omega_3(t,t_0) &=& \frac{i}{6} \int_{t_0}^{t}dt_1\int_{t_0}^{t_1} dt_2  \int_{t_0}^{t2} dt_3 \Big([H(t_1),[H(t_2),H(t_3)]]\nonumber\\ &+& [H(t_3),[H(t_2),H(t_1)]] \Big),\nonumber
\end{eqnarray}
and so on. In our case, $\Omega_k$ terms for $k>2$ are zero because $[H(s),[H(s'),H(s'')]]=0$. The first term $\Omega_1$ can be written as 
\begin{eqnarray}\label{Magnus1}
\Omega_1(t,t_0) =-i \sum_{j,m}\Delta_{jm}\big[a_m G_{jm}(t,t_0) +a_m^\dag G_{jm}^*(t,t_0)\big]\sigma_j^z
\end{eqnarray}
where 
\begin{eqnarray}\label{Gfunc}
G_{jm}(t,t_0) =\int_{t_0}^t dt' f_j(t')e^{-i\nu_m t'}.
\end{eqnarray}
The second term can be calculated to be
\begin{widetext}
\begin{eqnarray}\label{Magnus2}
\Omega_2(t,t_0) &=& -\frac{1}{2}\int_{t_0}^{t}dt_1\int_{t_0}^{t_1} dt_2 [H(t_1),H(t_2)] = -\frac{i}{2}\int_{t_0}^{t}dt_1 \big[H(t_1),\Omega_1(t_1,t_0)] \nonumber\\ 
&=& -\frac{i}{2}\int_{t_0}^{t}dt_1\sum_{jm}\sum_{j'm'}(-i)\Delta_{jm}\Delta_{j'm'}[f_j(a_me^{-i\nu_mt_1}+a_{m}^\dag e^{i\nu_m t_1})\sigma_j^z,(a_{m'}G_{j'm'}+a_{m'}^\dag G^*_{j'm'} )\sigma_{j'}^z] \nonumber\\
&=& -\frac{i}{2}\int_{t_0}^{t}dt_1\sum_{jj'}\sum_{m}(-i)\Delta_{jm}\Delta_{j'm}\big(f_jG^*_{j'm}e^{-i\nu_m t_1}[a_m,a_m^\dag]+ f_{j}G_{j'm}e^{i\nu_m t_1}[a_m^\dag,a_m]\big)\sigma_j^z\sigma^z_{j'} \\
&=& i\sum_m \int_{t_0}^{t}dt_1\frac{1}{2i}\Delta_{1m}\Delta_{2m}\big(f_1G_{2m}+f_2G_{1m}\big)e^{i\nu_m t_1}\sigma_1^z\sigma^z_{2} +{\rm H.c.} +K(t,t_0)\mathds{1},\nonumber
\end{eqnarray}
\end{widetext}
where, we have made use of  $[a_m,a_{m'}^\dag]=\delta_{m,m'}$ and $(\sigma_j^z)^2=\mathds{1}$. In a more convenient manner, $\Omega_2(t,t_0)$ can be expressed as
\begin{eqnarray}\label{Magnus22}
\Omega_2(t,t_0) &=& i\varphi(t,t_0)\sigma_1^z\sigma_2^z + K(t,t_0)\mathds{1},
\end{eqnarray}
where the phase $\varphi$ is a time dependent function given by 
\begin{widetext}
\begin{eqnarray}\label{GenPhase}
\varphi(t,t_0)=\sum_m \int_{t_0}^{t}dt_1\frac{1}{2i}\Delta_{1m}\Delta_{2m}\big\{f_1(t_1)G_{2m}(t_1,t_0)+f_2(t_1)G_{1m}(t_1,t_0)\big\}e^{i\nu_m t_1} +{\rm H.c.} ,
\end{eqnarray}
and we have ignored the term $K(t)$, as it will only contribute with a global phase. This can be equivalently written as 
\begin{eqnarray}\label{GenPhase2}
\varphi(t,t_0)=\sum_m  \Im{ \int_{t_0}^{t}dt_1\Delta_{1m}\Delta_{2m}\big\{f_1(t_1)G_{2m}(t_1,t_0)+f_2(t_1)G_{1m}(t_1,t_0)\big\}e^{i\nu_m t_1}} ,
\end{eqnarray}
\end{widetext}
where $\Im$ indicates the imaginary part.
Finally, we can easily check that the time evolution operator can be written as 
\begin{equation}\label{timevol}
U(t,t_0) =U_s(t,t_0)U_c(t,t_0) 
\end{equation}
where 
\begin{equation}\label{twogate}
U_s(t,t_0)= \exp \bigg\{-\! i \sum_{j,m}\Delta_{jm} \left[ a_m G_{jm}(t,t_0) + a_m^\dag G_{jm}^*(t,t_0)\right]\sigma_j^z \bigg\},
\end{equation}
and
\begin{equation}
U_c(t,t_0)=\exp \left[i \varphi(t,t_0) \sigma_1^z \sigma_2^z\right].
\end{equation}

\section*{Appendix E: Properties of the $G_{jm}(t)$ and $\varphi(t)$ functions}

\begin{figure*}[t!]
\centering
\includegraphics[width=1\linewidth]{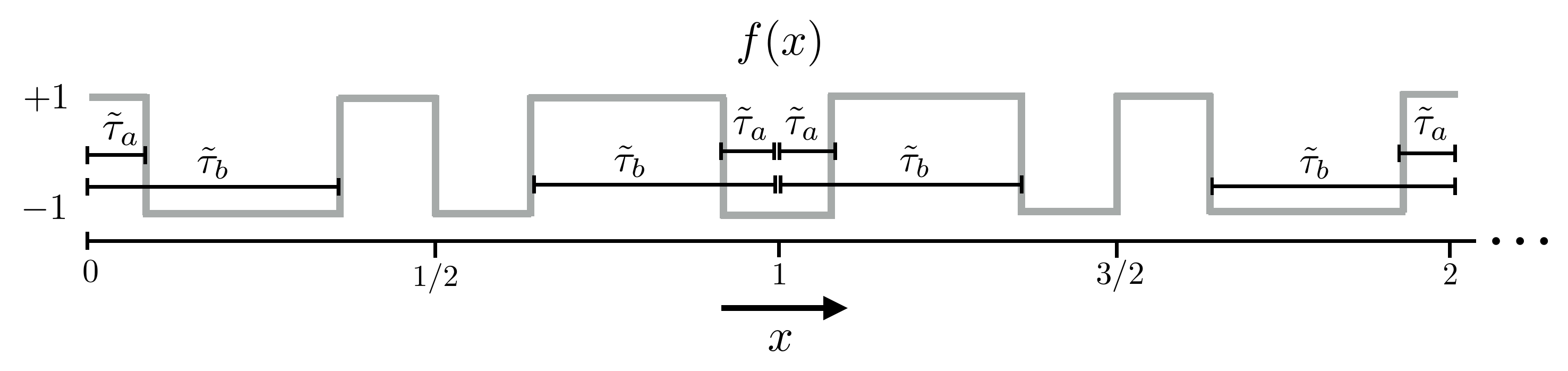}
\caption{ Modulation function $f(x)$ function that corresponds to the first two blocks of the AXY-4 pulse sequence. Time has been normalised by the characteristic time of the sequence $\tau$ ($x=t/\tau$), as well as $\tilde{\tau}_a=\tau_a/\tau$ and $\tilde{\tau}=\tau_b/\tau$.}\label{fig:AXY2}
\end{figure*}

Searching for all different sequences that fulfill the conditions $G_{jm}(T_f)=0$ and $\varphi(T_f)\neq0$ gets easy if we identify which are the indispensable variables that define the problem. The sequence function $f_1(t)=f_2(t)=f(t)$ is completely defined by the four parameters $\tau_a,\tau_b,\tau$, and $M$, for the case of an AXY-$M$ sequence. The duration of the sequence is of course only determined by two of them: $\tau$ and $M$ ($T_f=M\tau$). It is useful to rescale the domain on the $f$ function using $\tau$, the characteristic time of the sequence, as $t=x\tau$. Then, the property $f(t+\tau)=-f(t)$ becomes $f(x+1)=-f(x)$, and also $\tau_a,\tau_b$ may be redefined as $\tilde{\tau}_a=\tau_a/\tau$ and $\tilde{\tau}_b=\tau_b/\tau$. The AXY-4 sequence with this time rescaling is shown in Fig.~\ref{fig:AXY2}.  With this change of variable, the $G_{jm}$ functions at a time $T_f$ read
\begin{equation}\label{Gchange}
G_{jm}(T_f)=\int_{0}^{T_f}\!dt \ f(t)\ e^{-i\nu_m t}=\tau\int_{0}^{M}\!dx \ f(x) \ e^{-i\nu_m \tau x}.
\end{equation}
Now, if we relate $\tau$ and $\nu$ as
\begin{equation}\label{nutau}
\nu \tau=2\pi r \ \ \mbox{with} \ \ r\in \mathbb{N}, 
\end{equation}
we can construct functions that are independent of the frequency $\nu$ as
\begin{widetext}
\begin{eqnarray}\label{Gtilde}
\tilde{G}_{j1}(T_f)\equiv \nu_1G_{j1}(T_f)=\nu_1\tau \int_{0}^{M}\!dx \ f(x)\ e^{-i2\pi r x}=2\pi r\int_{0}^{M}\!dx \ f(x)\ e^{-i2\pi r x} \\
\tilde{G}_{j2}(T_f)\equiv \nu_2G_{j2}(T_f)=\nu_2\tau \int_{0}^{M}\!dx \ f(x)\ e^{-i2\pi\sqrt{3} r x}=2\pi \sqrt{3}r\int_{0}^{M}\!dx \ f(x)\ e^{-i2\pi\sqrt{3} r x}.
\end{eqnarray}
The same procedure can be followed for the $\varphi(t)$ function, that can be redefined in terms of $\tilde{\varphi}_1(t)$ and $\tilde{\varphi}_1(t)$ as 
\begin{eqnarray}\label{varphi}
\varphi(t)&=&\Bigg(\frac{\Delta_1}{\nu_1}\Bigg)^2\tilde{\varphi}_1(t)-\Bigg(\frac{\Delta_2}{\nu_2}\Bigg)^2\tilde{\varphi}_2(t)=\Bigg(\frac{\Delta_1}{\nu_1}\Bigg)^2(\tilde{\varphi}_1(t)-\frac{1}{3\!\sqrt{3}}\tilde{\varphi}_2(t))=\Bigg(\frac{\Delta_1}{\nu_1}\Bigg)^2\tilde{\varphi}(t),
\end{eqnarray}
where 
\begin{eqnarray}\label{phitilde}
\tilde{\varphi}_1(T_f)=(2\pi r)^2 \ \Im{\int_{0}^{M}\!dx \int_{0}^{x}\!dy \ [f_1(x)f_2(y)+f_2(x)f_1(y)] \ e^{i2\pi r(x-y)}} \\
\tilde{\varphi}_2(T_f)=(2\pi \sqrt{3}r)^2 \ \Im{ \int_{0}^{M}\! dx\int_{0}^{x}\!dy \ [f_1(x)f_2(y)+f_2(x)f_1(y)] \ e^{i2\pi \sqrt{3}r(x-y)}},
\end{eqnarray}
or
\begin{eqnarray}\label{phitilde}
\tilde{\varphi}(T_f)=(2\pi r)^2 \ \Im{\int_{0}^{M}\!dx \int_{0}^{x}\!dy \ [f_1(x)f_2(y)+f_2(x)f_1(y)] \ (e^{i2\pi r(x-y)}-\frac{1}{\sqrt{3}}e^{i2\pi \sqrt{3}r(x-y) })}. 
\end{eqnarray}
\end{widetext}
Now, it is clear that the functions $\tilde{G}_{j1}(T_f)$, $\tilde{G}_{j2}(T_f)$, and $\tilde{\varphi}(T_f)$, only depend on $\tilde{\tau}_a$, $\tilde{\tau}_b$, $M$ and $r$. Therefore, the functions plotted in Fig.~(3) of the main text (corresponding to cases $M=4$, $r=1,2,3$), do not depend on parameters $\nu_m$ and $\Delta_m$, but only on $\tilde{\tau}_a$ and $\tilde{\tau}_b$.

\section*{Appendix F: Pulse propagator}\label{PulseP}
\begin{figure}[t]
\centering
\includegraphics[width=1\linewidth]{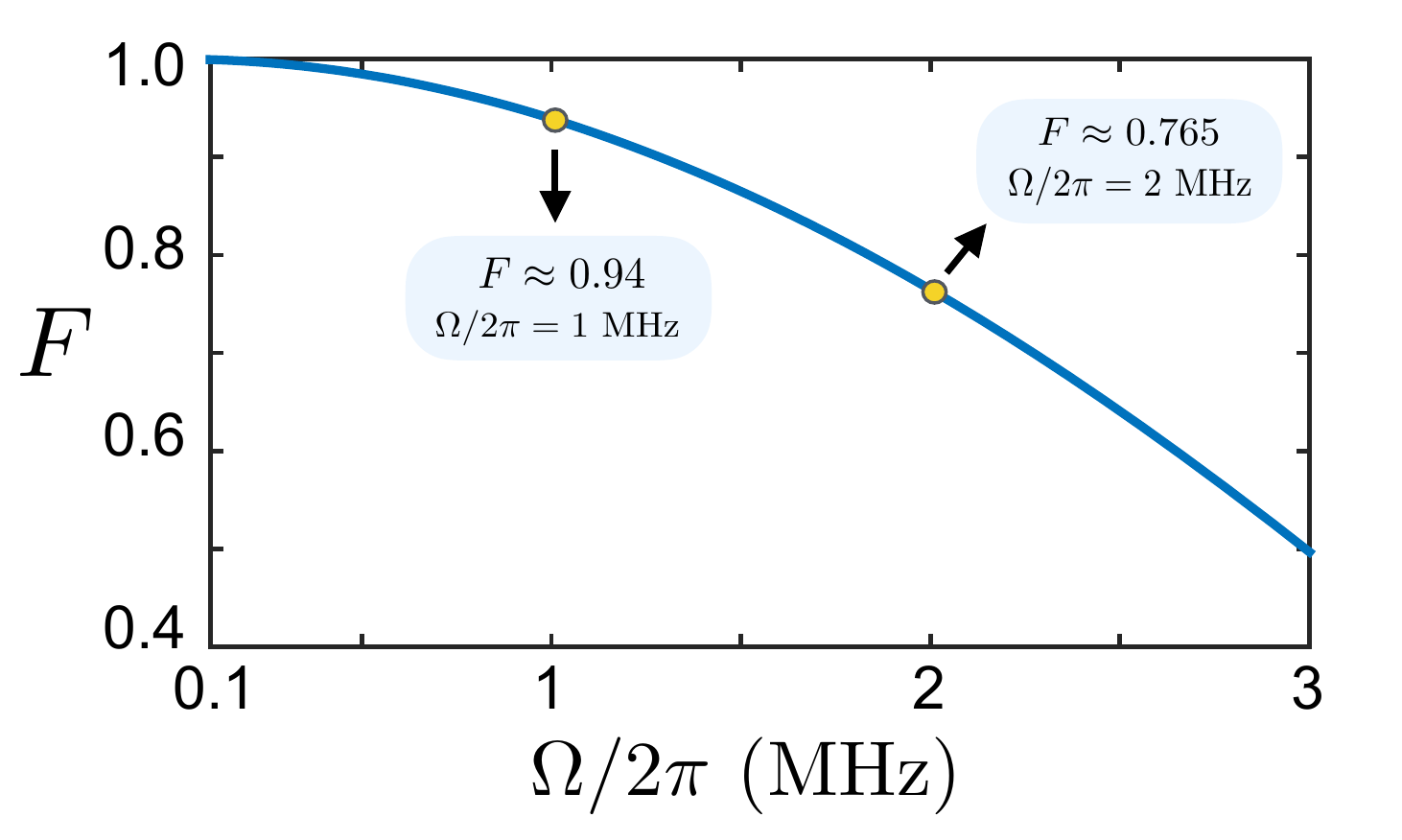}
\caption{Fidelity after 20 $\pi$-pulses between the propagators associated to the Hamiltonians $H =  \frac{\Omega_1}{2} \sigma_1^{\phi_1} +  \frac{\Omega_1}{2}[ \sigma_2^+  e^{i\phi_1} e^{i\delta_2t} + {\rm H.c.}]$ and $H =  \frac{\Omega_1}{2} \sigma_1^{\phi_1}$ as a function of the Rabi frequency $\Omega$.  We can observe how the fidelity decays because of the fail of the RWA. For this plot we have taken $\delta_2\approx (2\pi)\times 45$ MHz, a value that is even larger than the ones we will have our  method based on $\pi$-pulses gets displayed.}
\label{crosstalkfigure}
\end{figure}

Our strategy to produce fast $\pi$-pulses on a certain ion qubit and, at the same time, eliminate undesired effects on the off-resonant qubit consists in appropriately choosing the Rabi frequency of the driving. When a $\pi$-pulse is applied, say on the first qubit, we have the following Hamiltonian
\begin{equation}
H = \frac{\omega_1}{2} \sigma_1^z + \frac{\omega_2}{2} \sigma_2^z + \Omega_1 \cos(\omega_1 t - \phi_1) (\sigma_1^x + \sigma_2^x ).
\end{equation}
In a rotating frame with respect to $ \frac{\omega_1}{2} \sigma_1^z + \frac{\omega_2}{2} \sigma_2^z$ and after eliminating fast rotating terms, the above Hamiltonian reads
\begin{eqnarray}\label{ctH}
H =  \frac{\Omega_1}{2} \sigma_1^{\phi_1} +  \frac{\Omega_1}{2}[ \sigma_2^+  e^{i\phi_1} e^{i\delta_2t} + {\rm H.c.}],
\end{eqnarray}
where $\sigma_1^{\phi_1} =\sigma_1^+ e^{i\phi_1} + \sigma_1^- e^{-i\phi_1}$ and $\delta_2 = \omega_2 - \omega_1$. At this level one can argue that, only when the Rabi frequency is small, the second term on the r.h.s. of the above equation is negligible. This unavoidably limits the value of $\Omega_1$ and, consequently, how fast our decoupling pulses can be. In this respect, note that  for a value of $\omega_2 - \omega_1\approx (2\pi)\times45$ MHz, which is even larger than the ones used in the article, we find that $\Omega_1$ should be significantly smaller than $(2\pi)\times1$ MHz if we want to eliminate the crosstalk between ions during 20 $\pi$-pulses, see Fig.~\ref{crosstalkfigure}. 

To eliminate this restriction, we can use the following expression 
\begin{eqnarray}\label{prop}
U_{[t:t_0]} &\equiv& \hat{T}e^{-i\int_{t_0}^t H(s) ds} = U_0 \tilde{U}_{[t:t_0]} \\ \nonumber
&\equiv& e^{-iH_\delta (t-t_0)} \hat{T}e^{-i\int_{t_0}^t  U^{\dag}_0 (H(s) - H_\delta) U_0ds} ,
\end{eqnarray}
where $\hat{T}$ is the time ordering operator, $H_\delta=-(\delta_2/2) \sigma_2^z$, and find the time evolution operator for Eq.~(\ref{ctH}) in a generic time interval $(t, t_0)$. This is 
\begin{equation}
U_{[t:t_0]} =  e^{-i \frac{\Omega_1}{2} \sigma_1^{\phi_1} (t-t_0) } e^{i\frac{\delta_2}{2} \sigma_2^{z} (t-t_0)} e^{-i  \gamma (t-t_0)  \tilde\sigma_2} ,
\end{equation}
where $\gamma = \frac{1}{2}\sqrt{ \Omega_1^2 +  \delta_2^2 }$,
 \begin{equation}
 \tilde\sigma_2 = \frac{\Omega_1}{2 \gamma} e^{i\frac{\delta_2}{2}t_0\sigma_2^z} \sigma_2^{\phi_1}e^{-i\frac{\delta_2}{2}t_0\sigma_2^z}  + \frac{\delta_2}{2 \gamma} \sigma_2^{z}.
 \end{equation}

Note that the first and the second terms in Eq.~(\ref{ctH}) commute, which allows to apply the relation~(\ref{prop}) only to the part $\frac{\Omega_1}{2}[ \sigma_2^+  e^{i\phi_1} e^{i\delta_2t} + {\rm H.c.} ]$. Finally, for the sake of realising a $\pi$-pulse we will set $(t-t_0) = t^{(1)}_{\pi} \equiv \frac{\pi}{\Omega_1}$ which gives rise to 
\begin{eqnarray}\label{piuno}
U_{t_\pi}^{(1)} =  e^{-i \frac{\Omega_1}{2} \sigma_1^{\phi_1} t_\pi } e^{i\frac{\delta_2}{2} \sigma_2^{z} t_\pi} e^{-i  \gamma t_\pi \tilde\sigma_2}. 
\end{eqnarray}

In the same manner, for a $\pi$-pulse (with $ t^{(2)}_{\pi} \equiv \frac{\pi}{\Omega_2}$) in resonance with the second ion we would have 
\begin{eqnarray}\label{pidos}
U_{t_\pi}^{(2)} =  e^{-i \frac{\Omega_2}{2} \sigma_2^{\phi_2} t_\pi } e^{i\frac{\delta_1}{2} \sigma_1^{z} t_\pi} e^{-i  \gamma t_\pi  \tilde\sigma_1}. 
\end{eqnarray}
Equations~(\ref{piuno}) and~(\ref{pidos})  contain the $\pi$-pulses in which we are interested ($e^{-i \frac{\Omega_1}{2} \sigma_1^{\phi_1} t_\pi }$ and $e^{-i \frac{\Omega_2}{2} \sigma_2^{\phi_2} t_\pi }$) plus the crosstalk contributions we want to get rid off. To eliminate terms $e^{-i  \gamma t_\pi \tilde\sigma_2}$ and $e^{-i  \gamma t_\pi  \tilde\sigma_1}$, we will adjust the Rabi frequencies $\Omega_{1,2}$ such that 
\begin{equation}
\gamma t_\pi = \frac{1}{2} \sqrt{ (\Omega_{1,2})^2 +  (\delta_{2,1})^2 }\frac{\pi}{\Omega_{1,2}}   = \pi \times k, \mbox{with} \ k\in \mathbb{Z}. 
\end{equation}
In this case we have that $e^{-i  \gamma t_\pi \tilde\sigma_2} = e^{-i  \gamma t_\pi  \tilde\sigma_1} = \pm \mathds{1}$, i.e. the unwanted terms contribute as a global phase. Hence, only pure dephasing terms remain in both pulses, $e^{i\frac{\delta_2}{2} \sigma_2^{z} t_\pi}$ and $e^{i\frac{\delta_1}{2} \sigma_1^{z} t_\pi}$, which will be cancelled by our tailored microwave sequences as explained in the main text. 

\section*{Appendix G: Heating rates estimation}\label{estimation}
To estimate the $\Gamma_{b,c}$ parameters, we will rely on the data provided in~\cite{Weidt16} and by~\cite{Hensinger17}, as well as on the scaling relations one can extract from~\cite{Brownnutt15}. More specifically we take as reference values (for the com mode) $\dot{n}_{\rm com}^{\rm ref} = 41$ phonons/second for a frequency $\nu^{\rm ref}_1/(2\pi) = 427$ kHz, and (for the breathing mode) $\dot{n}_{\rm bre}^{\rm ref} = 7$ phonons/second for a frequency $\nu^{\rm ref}_2/(2\pi) = 459$ kHz,~\cite{Weidt16, Hensinger17}. The latter values correspond to a configuration at room temperature ($T^{\rm ref}=300$ K) with an ion-electrode distance of $d^{\rm ref} \approx 310 \ \mu$m, that would give rise to a magnetic field gradient $g_B = 23.6 {\rm \frac{T}{\rm m}}$. 

Our operating conditions require, for the first studied case, an ion-electrode distance of $d\approx 150 \ \mu$m, to generate a magnetic field gradient of $g_B = 150 {\rm \frac{T}{\rm m}}$ where $\nu_1=\nu$ and $\nu_2 = \sqrt{3} \nu$ with $\nu/(2\pi) = 150$ kHz, while we will consider low temperatures of $T=50$ K. In this situation one can derive new values for $\dot{n}_{\rm com}$ and $\dot{n}_{\rm st}$ using scaling relations~\cite{Brownnutt15} which in our case are
\begin{eqnarray}\label{scaling1}
\dot{n}_{\rm com}\approx \dot{n}_{\rm com}^{\rm ref} \ \bigg(\frac{\nu_1^{\rm ref}}{\nu_1}\bigg)^2 \bigg(\frac{d^{\rm ref}}{d}\bigg)^4 \bigg(\frac{T^{\rm ref}}{T}\bigg)^{-2.13},
\end{eqnarray}
and
\begin{eqnarray}\label{scaling2}
\dot{n}_{\rm bre}\approx \dot{n}_{\rm bre}^{\rm ref} \ \bigg(\frac{\nu_2^{\rm ref}}{\nu_2}\bigg)^2 \bigg(\frac{d^{\rm ref}}{d}\bigg)^4 \bigg(\frac{T^{\rm ref}}{T}\bigg)^{-2.13}.
\end{eqnarray}
Then, one can use that, when close to the motional ground state, we have~\cite{Brownnutt15}  
\begin{equation}
\dot{n}_{\rm com, bre} = \Gamma_{b,c} \ \bar{N}_{b, c},
\end{equation}
that together with the definition of $\bar{N}_{b, c} \equiv N^{\rm thermal}_{b,c} =1/(e^{\hbar \nu_{1,2}/k_{\rm B} T}-1)$, allows us to obtain the values for $\Gamma_{b,c}$. 

In the second studied case, a magnetic field gradient of $g_B = 300 \frac{\rm T}{\rm m}$ would require to locate the ions closer to the electrodes, which would induce more heating. We estimate a distance according to the relation $d = \sqrt{\frac{150}{300}} \ 150 \ \mu$m $\approx 106 \ \mu$m that assumes a dependence  $g_B \sim \frac{1}{d^2}$. This new distance can be used in Eqs.~(\ref{scaling1}) and~(\ref{scaling2}) to derive new values for the heating rates $\Gamma_{b,c}$.

\end{document}